\documentclass{scipost}

\usepackage{graphicx}
\usepackage{amsmath}
\usepackage{bm}
\usepackage{siunitx}
\usepackage{hyperref}
\usepackage{color}
\usepackage{braket}
\usepackage{soul}
\soulregister\cite7
\usepackage[normalem]{ulem}
\usepackage[title]{appendix}

\hypersetup{
    colorlinks,
    linkcolor={red!50!black},
    citecolor={blue!50!black},
    urlcolor={blue!80!black}
}

\usepackage[bitstream-charter]{mathdesign}
\DeclareSymbolFont{usualmathcal}{OMS}{cmsy}{m}{n}
\DeclareSymbolFontAlphabet{\mathcal}{usualmathcal}

\fancypagestyle{SPstyle}{
\fancyhf{}
\lhead{\colorbox{scipostblue}{\bf \color{white} ~SciPost Physics }}
\rhead{{\bf \color{scipostdeepblue} ~Submission }}

\fancyfoot[C]{\textbf{\thepage}}
}

\DeclareMathOperator{\sgn}{sgn}

\usepackage[most]{tcolorbox}
\usepackage{xcolor}

\newtcolorbox[auto counter, number within=section]{comentario}[1][]{
    colback=blue!5!white,
    colframe=blue!75!black,
    title={Lemma~\thetcbcounter:~Sufficient conditions, albeit not necessary},
    #1
}

\newtcolorbox[use counter from=comentario]{comentario2}[1][]{
    colback=blue!5!white,
    colframe=blue!75!black,
    title={Lemma~\thetcbcounter:~Necessary conditions},
    #1
}

\begin{document}

\pagestyle{SPstyle}

\begin{tikzpicture}[remember picture,overlay]
  \node[anchor=north east, xshift=-1.5cm, yshift=-1cm] at (current page.north east) {\textsf{IPARCOS-UCM-26-056}};
\end{tikzpicture}

\begin{center}
{\Large \textbf{\color{scipostdeepblue}{
Theoretical constraints on beyond Standard Model extensions of the Higgs potential
}}}
\end{center}

\begin{center}\textbf{
Javier E. Pena-Vico\textsuperscript{1$\dagger$}, 
Carlos Quezada-Calonge\textsuperscript{2$\S$} and
Juan J. Sanz-Cillero\textsuperscript{1$\star$}
}\end{center}

\begin{center}

{\bf 1} Dept. Física Teórica and IPARCOS, Universidad Complutense de Madrid,
Plaza de las Ciencias 1, 28040 Madrid, Spain

{\bf 2} Dipartimento di Fisica, Università di Torino, and INFN, Sezione di Torino,
Via P. Giuria 1, 10125 Torino, Italy

$\dagger$ \href{mailto:jadelape@ucm.es}{\small jadelape@ucm.es}\ ,\quad 
$\S$ \href{mailto:carlosariel.quezadacalonge@unito.it}{\small carlosariel.quezadacalonge@unito.it}\ ,\quad 
$\star$ \href{mailto:jjsanzcillero@ucm.es}{\small jjsanzcillero@ucm.es}

\end{center}

\section*{\color{scipostdeepblue}{Abstract}}
\textbf{\boldmath{
A fundamental requirement for the consistency of quantum field theories is that their Higgs potential is lower-bounded, such that no direction in the $\Phi$ field space leads to $V(\Phi)\to \, -\infty$. This work presents a comprehensive analysis of the various theoretical constraints on potentials in extended Higgs sectors, highlighting their relative impact and restrictiveness. We illustrate our study on the potential parameters with two particular Higgs sector extensions: the Real Singlet Extension (RSE) and the Two-Higgs-Doublet Model (2HDM), separating $\mathbb{Z}_{2}$-conserving and $\mathbb{Z}_{2}$-violating cases. In order to do this, we derive and compare these conditions to other studies found in the literature. Additionally, we developed a publicly available code capable of testing positivity conditions in complex potentials, including particularly challenging cases such as the general $\mathbb{Z}_2$-violating 2HDM. Finally, we apply those restrictions to the physical parameters and deduce an upper and a lower bound on the masses of new heavy scalars.
}}

\vspace{\baselineskip}

\noindent\textcolor{white!90!black}{
\fbox{\parbox{0.975\linewidth}{%
\textcolor{white!40!black}{\begin{tabular}{lr}%
  \begin{minipage}{0.6\textwidth}%
    {\small Copyright attribution to authors. \newline
    This work is a submission to SciPost Physics. \newline
    License information to appear upon publication. \newline
    Publication information to appear upon publication.}
  \end{minipage} & \begin{minipage}{0.4\textwidth}
    {\small Received Date \newline Accepted Date \newline Published Date}%
  \end{minipage}
\end{tabular}}
}}
}


\vspace{10pt}
\noindent\rule{\textwidth}{1pt}
\tableofcontents
\noindent\rule{\textwidth}{1pt}
\vspace{10pt}

\section{Introduction}

Extensions of the Standard Model (SM) Higgs sector through the inclusion of additional scalars constitute a popular approach to beyond Standard Model (BSM) phenomenological studies. 
While the SM Higgs sector includes only one complex doublet, which can be understood as four real scalar fields, extended models introduce additional degrees of freedom by enlarging the number of scalar fields in the Higgs sector. Some of the most popular renormalizable extensions
studied within the literature are the Real and Complex Singlet extensions~\cite{Robens_2015,Bosse:2026bdk,Buchalla_2017,Bosse:2026rni,Kannike_2016,kannike2024rebuttalnotevacuumstability,BISWAS_2013,Barger_2009,Cho_2023} and the Two-Higgs-Doublet~Model~(2HDM)~\cite{Branco_2012,Davidson:2005cw,Ginzburg_2005,Ginzburg_2005_2,ginzburg2005symmetries2hdmdifferentvacua,Gunion_2003,kastening1993boundsstabilitysymmetrybreaking,Deshpande:1977rw,Dawson_2022,Baum_2018,CilleroQuezadaHiggsbasis,gonzalez-jerez-quezada-sanz-cillero-2017,Boto_2020,ginzburg2003treelevelunitarityconstraints2hdm,Dawson:2023oce,Faro:2020qyp,Arroyo-Urena:2026vgf,Alakhras:2017bbe,Kannike:2012pe,Maniatis:2006fs,Klimenko:1984qx,Nie:1998yn,Kanemura:1999xf,Dittmaier:2026nnb,Chakrabortty:2013mha,Atwood_1997,Camacho:2026qsr,Arco:2022jrt,Arco:2023sac,Matuszak:2026xsz,Moretti:2026qsw}.

Studies on these Higgs extensions\footnote{We will adopt the name ``Higgs extensions'' to refer to a broad class of models that extend the Standard Model by adding new multiplets.} typically analyze the positivity of the potential in the limit when the fields tend to infinity~\cite{Gunion_2003,kastening1993boundsstabilitysymmetrybreaking,Branco_2012,Kannike:2012pe,Maniatis:2006fs,Klimenko:1984qx,Nie:1998yn,Chakrabortty:2013mha,Ivanov:2018jmz}. 
These works impose that the potential increases as the fields grow and are lower bounded, often referred generically as ``positivity restrictions''. This condition consequently constrains the valid values of the parameters of the Higgs extension. Often, for the models considered in this work, the constraints obtained by requiring positivity in some studies \cite{Robens_2015,Buchalla_2017,Bosse:2026bdk,Bosse:2026rni} are too
restrictive, since they disregard some parameter configurations that correspond to potentials that still obey positivity.

This idea is not novel and has been very prolific in the literature. For example, for the Real and Complex Singlet extensions, the copositivity conditions can be found in Refs.~\cite{Kannike:2012pe,Robens_2015,Kannike_2016,kannike2024rebuttalnotevacuumstability,Barger_2009}, and for the $\mathbb{Z}_2$-conserving 2HDM in Refs.~\cite{Deshpande:1977rw,Ginzburg_2005,Maniatis:2006fs,Kannike:2012pe,Kannike_2016,Klimenko:1984qx,Nie:1998yn,Kanemura:1999xf}. Furthermore, some $\mathbb{Z}_2$-violating extensions that are not the most general 2HDM can be found in Ref.~\cite{Maniatis:2006fs}. In addition to this, we aim to find the copositivity restrictions for the most general $\mathbb{Z}_2$-violating 2HDM with a computational approach. Note that this viewpoint is not new since some numerical methods have been used in other works to study certain Higgs extensions~\cite{Ivanov:2018jmz,Wojcik:2023usm,Boto:2022uwv,Jurciukonis:2026dmc}; however, as far as we are concerned, no similar results can be found in the literature.

This paper is structured as follows. In Sec.~\ref{sec:positivity} we introduce the conditions which will be used in order to assess whether a field configuration conforms to positivity, meaning it is bounded from below. In Sec.~\ref{section2}, we consider a Real Singlet Extension (RSE) of the Standard Model and we recover the strictest and the most relaxed conditions already given in Refs.~\cite{Gunion_2003,Kannike_2016} and comment on them. In Sec.~\ref{section3}, we study in detail the general 2HDM: first, we start with a $\mathbb{Z}_2$-symmetry conserving scenario, and progressively include the remaining 2HDM terms that violate the $\mathbb{Z}_2$ symmetry. 
In Subsec. \ref{subsect_softlybrokenZ2}, we compare the positivity conditions we derive for the case with $\mathbb{Z}_2$-symmetry to the restrictions found in Refs.~\cite{Deshpande:1977rw,ginzburg2005symmetries2hdmdifferentvacua, Gunion_2003}. 
In Subsec.~\ref{section4}, we derive analytical restrictions for two interesting subcases of the general 2HDM Lagrangian.
Furthermore, these results are corroborated through a numerical algorithm we developed, a {\it Rust} code that minimizes the potential, publicly available in Ref.~\cite{javi}.
This code has also been later employed to numerically constrain the parameters of the potential of the general 2HDM. We stress that this is a novel result and no similar outcome has been provided in previous literature, to the best of our knowledge.
In Sec.~\ref{sec:pheno-2HDM-softly-broken}, we perform a phenomenological study of the impact of the constraints obtained for the 2HDM with a softly-broken $\mathbb{Z}_{2}$ symmetry in terms of physical parameters. In particular, we show how the constraints yield upper and lower bounds on the heavy BSM scalar masses.  
Finally, we summarize our findings and conclusions in Sec.~\ref{sec:conclusions}. Some technical details have been relegated to the Appendices. 

\section{Positivity conditions on the potential}\label{sec:positivity}

Regardless of which Higgs extension is being studied, a traditional requirement in quantum field theories is to have a potential such that no direction in the $\Phi$ field space leads to $V(\Phi)\to \, -\infty$, herein referred as ``positivity restrictions'' or bounded from below (BFB). Given a convenient choice of the potential energy origin, studying the positivity of the potential of a new theory is an appropriate form of assessing whether it is aligned with our modern and usual understanding of Quantum Field Theory (as the Standard Model does). For this reason, the positivity conditions of the Higgs potential constitute a cornerstone in the study of extensions of the Standard Model. In the following, we will state and discuss the conditions  to study the positivity of the potential for different constraints that we have named: ``sufficient constraints'' and ``necessary constraints''\footnote{In any case, we will be neglecting the quadratic terms of the potential for this discussion. A comment regarding these quadratic terms can be found in App.~\ref{quadratictermsandloops}.}.

In some studies \cite{Robens_2015,Buchalla_2017,Dittmaier:2026nnb}, the positivity restrictions are usually obtained by observing that for large-field configurations a renormalizable potential $V$ is dominated by its quartic term. One can then write     
the scalar field potential as the product
\begin{equation}
    V \sim\sum_{\alpha=[ij], \beta=[k\ell] } \Lambda_{\alpha\beta } \bigg(\phi^{i\, \dagger} \phi^j\bigg)\, \bigg( \phi^{k\, \dagger} \phi^\ell\bigg) = \Lambda_{\alpha\beta}\upsilon^{\alpha}\upsilon^{\beta}\ \ \ ,\ \ \ \mbox{where}\ \ \ \upsilon\sim\phi^{\dagger}\phi,
\end{equation}
and impose the quadratic form, $\Lambda$, to be positive definite. This restriction ensures that the potential will tend to infinity as fields, $\Phi$, grow, even if the region is not physically valid. We can state these constraints in the Lemma~\ref{comment:lemma-suff}, 
which is also often named as \textit{positivity}. 

\begin{comentario}[label={comment:lemma-suff}]
Let $\Lambda$ be a positive definite $n\times n$ matrix. By Sylvester's criterion, this is equivalent to requiring that all its leading principal minors are strictly positive. Then, the quadratic form
\[
V \equiv \Lambda_{\alpha\beta}\upsilon^{\alpha}\upsilon^{\beta}
\]
is always greater than zero for all nonzero vectors $\upsilon\sim\phi^{\dagger}\phi\in\mathbb{C}^{n}$.
\end{comentario}

The aim of this work is to study whether this condition overly restricts the parameter space, and how it is possible to find less strict conditions by allowing the potential to be unbounded from below in regions that are physically invalid. 
For the purpose of this work, the region $\|\Phi\|^2 < 0$ is regarded as physically invalid. Hence, the previous requirement can be softened into the Lemma \ref{comment:lemma-necessary}, which can also be stated in terms of \textit{copositivity} \cite{Chakrabortty:2013mha,Kannike:2012pe,Kannike_2016}.

\begin{comentario2}[label={comment:lemma-necessary}]
Given a quadratic form, $\Lambda_{n\times n}$ , that might not be positive definite, the scalar $$V \equiv \Lambda_{\alpha\beta}\upsilon^{\alpha}\upsilon^{\beta}$$ is always greater than zero for all nonzero vectors that live in the physical space (e.g.,~$\upsilon$~with~$\phi_j^{\dagger}\phi_j \geq  0$).
\end{comentario2}

Having stated both constraints in Lemmas \ref{comment:lemma-suff} and \ref{comment:lemma-necessary}, let us begin the study of the consequences of these conditions on the aforementioned BSM extensions. Therein, we will detail how $\upsilon^\alpha$ is constrained in the various models when restricting the analysis to physical $\phi_j$ configurations.

\section{Real Singlet Extension (RSE)}
\label{section2}
\label{sec:RSE}

We start by considering the simplest extension: an additional real singlet scalar, $S$. In this model, the potential term of the Lagrangian can be written as \cite{Buchalla_2017, Robens_2015, Dittmaier:2026nnb,Bosse:2026bdk,Bosse:2026rni,BISWAS_2013}:
\begin{equation}\label{DSHM_potential}
        V^{\mathbb{Z}_2}_{\si{RSE}}=-\frac{\mu_1^2}{2}\Phi^{\dagger}\Phi-\frac{\mu_2^2}{2}S^2+\frac{\lambda_1}{2}\left(\Phi^{\dagger}\Phi\right)^2
        +\frac{\lambda_2}{2}S^4+\lambda_3\Phi^{\dagger}\Phi S^2\ .
\end{equation}

To study the limit where the fields of the potential tend to infinity, the quadratic terms will be neglected, as usual, due to the dominance of quartic terms: 
\begin{equation}\label{quartictermssinglet}
    V^{\mathbb{Z}_2\ (4)}_{\si{RSE}}=\frac{1}{ 2
    }\begin{pmatrix}
        \Phi^{\dagger}\Phi,\ S^2
    \end{pmatrix}\underbrace{\begin{pmatrix}
        \lambda_1 & \lambda_3\\
        \lambda_3 & \lambda_2
    \end{pmatrix}}_{\Lambda}\begin{pmatrix}
        \Phi^{\dagger}\Phi\\
        S^2
    \end{pmatrix}\ .
\end{equation}

In order to bound the potential from below, it is first necessary to impose $\displaystyle\lim_{\Phi\to\infty}V(\Phi,0) > 0$ and $\displaystyle\lim_{S\to\infty}V(0,S) > 0$, which translates to
\begin{equation}
    \lambda_{1},\, \lambda_{2}>0\ .
\end{equation}

When the time comes to constraining $\lambda_{3}$, some studies impose that $\Lambda$ must be positive definite~\cite{Robens_2015,Buchalla_2017,Bosse:2026bdk,Bosse:2026rni}, according to the sufficient conditions of Lemma.~\ref{comment:lemma-suff}. However, as we will explictly see, this is too restrictive a condition as it implies that the potential tends to infinity in each quadrant. In order to find the actual necessary restrictions ---which are more relaxed--- given by Lemma~\ref{comment:lemma-necessary}, it is advisable to define
\begin{equation}\label{first_changeofvariable}
    x\equiv\Phi^{\dagger}\Phi\sqrt{\lambda_{1}}\ \mbox{and}\ y\equiv S^2\sqrt{\lambda_{2}}\, , 
\end{equation}
which allows us to rewrite Eq.~\eqref{quartictermssinglet} as:
\begin{equation}\label{DSHMpotential_newparameters}
    V^{\mathbb{Z}_2\ (4)}_{\si{RSE}}=\frac{1}{2}x^2+\frac{1}{2}y^2+\rho_3 xy\ ,
\end{equation}
where 
\begin{equation}
    \rho_{3}\equiv\frac{\lambda_{3}}{\sqrt{\lambda_{1}\lambda_{2}}}\ .
\end{equation}

Note that $x$ and $y$ are always positive. Hence, only the first quadrant contains physically relevant information of the potential. We must restrict ourselves to this quadrant.

\subsection{Necessary constraints on the quartic terms: quadratic field coordinates \texorpdfstring{$(x,y)$}{(x,y)}}

\begin{figure}[p]
    \centering
    \includegraphics[scale=0.575]{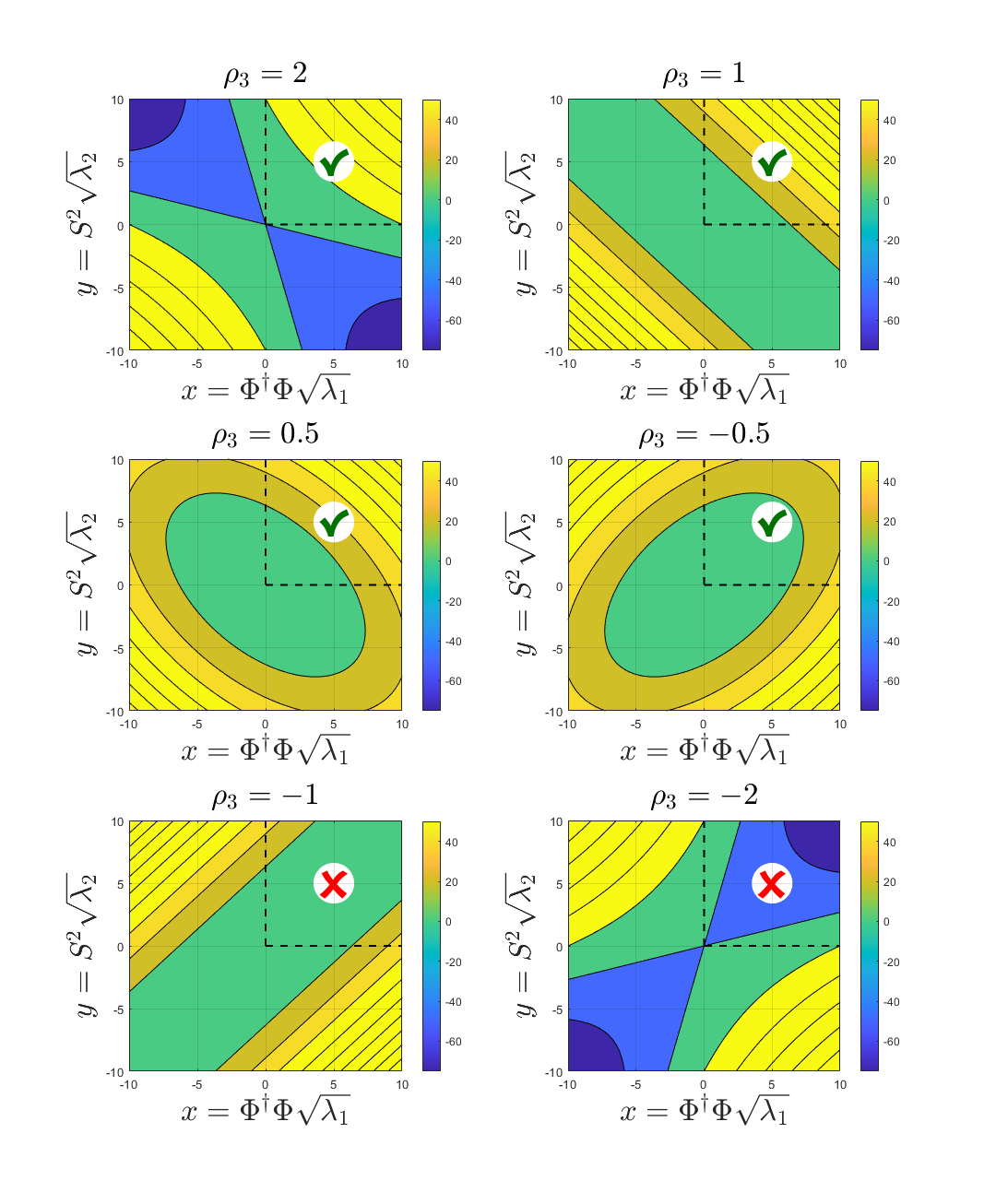} 
    \caption{Potential $V^{\mathbb{Z}_2}_{\si{RSE}}$ written in $x$ and $y$ variables and plotted for different values of $\rho_3$. Only those that obey Equation~\eqref{rho3gt-1} are valid.}
    \label{DSHM}
\end{figure}

We shall explore the quadrant completely by using straight lines ---parametrized as $y=x\tan\theta$. Furthermore, these coordinates will be useful when the quartic terms of the potential have a $\mathbb{Z}_{2}$ symmetry. For those Higgs extensions that do not present a $\mathbb{Z}_2$ symmetry, refer to Subsection \ref{polar}. For this model, the potential of the Real Singlet Extension conserves this symmetry and, therefore, it can be studied using this approach. In light of the foregoing results, the potential of this theory can be re-expressed in the new coordinates as
\begin{equation}\label{DSHM_reduced}
    V^{\mathbb{Z}_2\ (4)}_{\si{RSE}}=\frac{f(\theta)}{2}R^2\ \ ,\ \ \mbox{where}\ \ f(\theta)=1+\rho_3\sin(2\theta)\ \ \ \mbox{and}\ \ \ x=R\cos\theta,\, y=R\sin\theta\ .
\end{equation}

As stated previously in Lemma.~\ref{comment:lemma-suff} it is possible to find some positive restrictions by imposing $\Lambda$ to be positive definite. Nonetheless, there can exist some eigenvectors that live in the second and fourth quadrants whose eigenvalue is negative, which is not a physical region. Thus, focusing on finding the necessary conditions given the result explained in 
Lemma~\ref{comment:lemma-necessary}, we must restrict ourselves to the first quadrant, $\theta\in(0,\pi/2)$.

After imposing positivity, $V>0$, it is straightforward to derive that
\begin{equation}
    \rho_{3} > \frac{-1}{\sin{(2\theta)}},
\end{equation}
and the worst-case scenario occurs when $\sin{(2\theta)}$ takes its maximum value. As a consequence, we obtain that the minimal constraint to the quartic terms that is required to ascertain that the potential is bounded from below in the physical regions is \cite{Kannike:2012pe,Kannike_2016,kannike2024rebuttalnotevacuumstability}
\begin{equation}\label{rho3gt-1}
    \rho_{3}>-1.
\end{equation}
As it is shown in Fig.~\ref{DSHM}, those combinations of the parameters $\lambda_{1},\ \lambda_{2}$ and $\lambda_{3}$ that do not obey the necessary condition of Eq.~\eqref{rho3gt-1} are immediately discarded. Note that $\rho_{3}=-1$ leads us with a flat region as the worst scenario, which can be considered as valid since $V(\Phi)\nrightarrow \, -\infty$. However, this limit case will be discussed in the Appendix \ref{quadratictermsandloops} and will be marked as ``not valid'' hereinafter.

\subsection{Necessary constraints on the quartic terms: linear field coordinates \texorpdfstring{$(\sqrt{x},\sqrt{y})$}{(sqrt{x},sqrt{y})}}\label{polar}

\begin{figure}[p]
    \centering
    \includegraphics[scale=0.575]{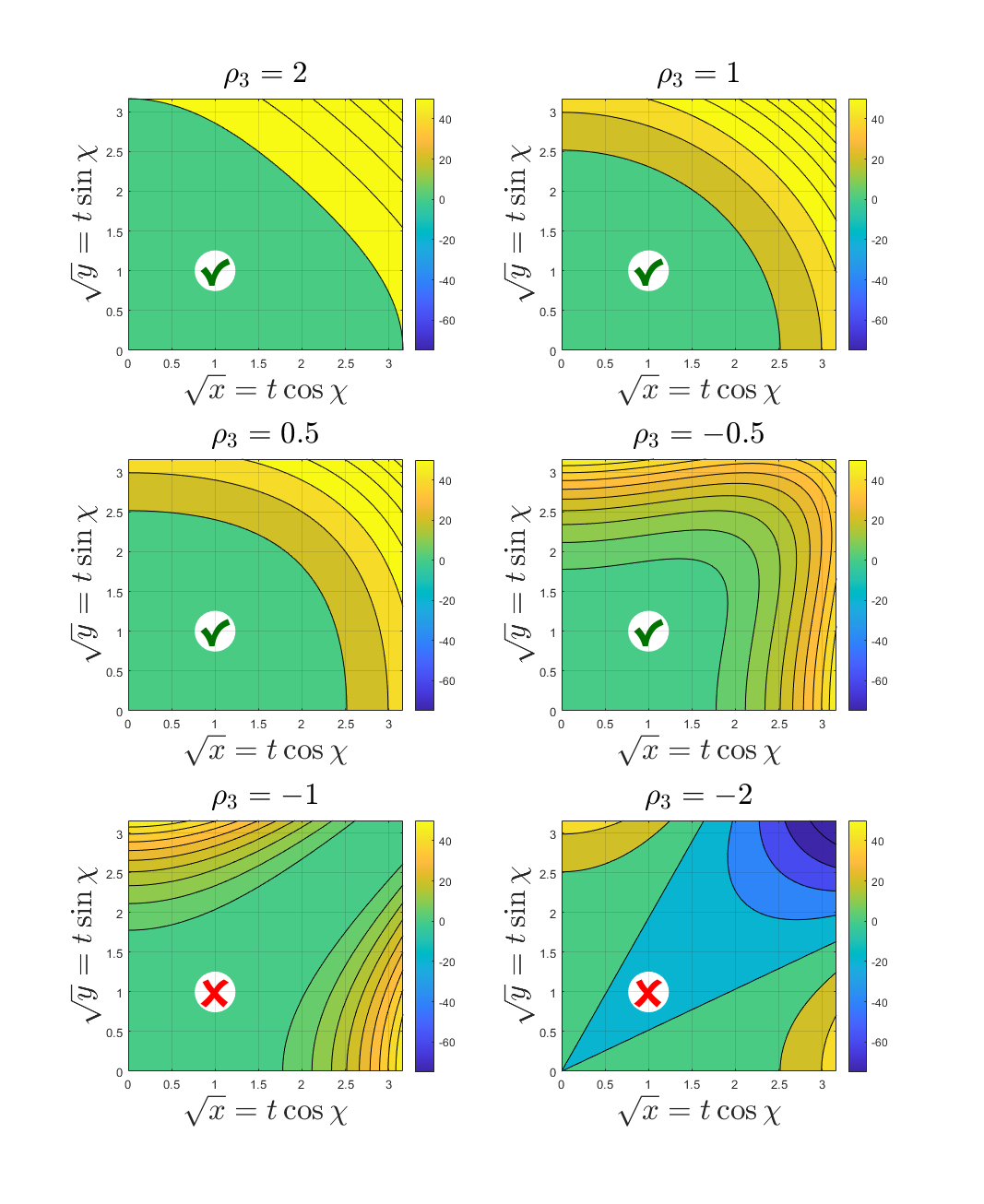} 
    \caption{Potential $V^{\mathbb{Z}_2}_{\si{RSE}}$ written in $\sqrt{x}$ and $\sqrt{y}$ variables and plotted for different values of $\rho_3$. Only those that obey Equation~\eqref{conditionsinglet} are valid. Note that there exists a relation between the coordinates used in Fig.~\ref{DSHM} and the ones used in this plot.}
    \label{DSHM_2}
\end{figure}

In this approach, we consider this change of variables
\begin{equation}\label{ourpolarcoordinates}
\begin{split}
    &\sqrt{x}=t\cos{\chi}\\
    &\sqrt{y}=t\sin{\chi}\ 
\end{split}
\ \ \ ,\ \ \ \chi\in\left(0,\frac{\pi}{2}\right).
\end{equation}
These coordinates considerably simplify the derivation of the constraints when the potential does not conserve the $\mathbb{Z}_2$ symmetry and, therefore, they will be used ---or at least a specific variant of them--- in some extensions of the Standard Model that will be studied later in this paper. Substituting these variables into Eq.~\eqref{DSHMpotential_newparameters} gives:
\begin{equation}
    V^{\mathbb{Z}_2\ (4)}_{\si{RSE}}=\frac{t^4}{2}\left[1+\frac{\left(\rho_{3}-1\right)}{2}\sin^2{\left(2\chi\right)}\right]\ , 
\end{equation}
where some half-angle identities have been used. Note that the radial term will always be positive. Since we want a potential that is bounded from below, we must impose that, for a certain radius, $V^{\mathbb{Z}_2\ (4)}_{\si{RSE}}>0$ in the first quadrant (first orthant), otherwise, the radial term will propagate a negative value in a certain direction of the first quadrant and the potential will not be bounded from below. After imposing this, one can derive the following:
\begin{equation}\label{sameingeneralpot}
    \rho_{3}-1>\frac{-2}{\sin^2{\left(2\chi\right)}}\ ,
\end{equation}
where the worst-case scenario occurs for $\sin^2{\left(2\chi\right)}=1$, which implies that \cite{Kannike:2012pe,Kannike_2016,kannike2024rebuttalnotevacuumstability}
\begin{equation}\label{conditionsinglet}
    \rho_3>-1\ .
\end{equation}
It is easy to observe in Fig.~\ref{DSHM_2} that imposing $\Lambda$ to be positive definite ---as the traditional sufficient constraints require--- narrows down the $\rho_{3}$ validity range to $(-1,1)$, while Equation~\eqref{conditionsinglet} is more permissive as it allows $\rho_{3}\in (-1,\infty)$.

Finally, it is important to point out that the previous results can be applicable to other theories whose potential also contains dominant quartic terms with the following structure:
\begin{equation}\label{structure}
    V \xrightarrow{x,y\rightarrow\infty} ax^2+by^2+cxy\ .
\end{equation}
Some examples of this are the models with a real or complex singlet regardless whether they carry a $\mathbb{Z}_2$ symmetry or not ---see \cite{BISWAS_2013,Barger_2009,Cho_2023}. This mathematical structure is also applicable when there is a weak breaking of $\mathbb{Z}_2$ symmetry only in the quadratic terms, provided that a series of details, which will be specified hereinafter, are taken into consideration.

\section{Two-Higgs Doublet Model (2HDM) potential}\label{section3}

Let us apply the previous conditions to a more involved model. The Two-Higgs Doublet Model (2HDM) extends the SM by considering an extra complex doublet. The most general 2HDM potential can be written as ---considering only terms of lower order than five \cite{Gunion_2003,Ginzburg_2005_2,ginzburg2003treelevelunitarityconstraints2hdm,Boto_2020,Ginzburg_2005,ginzburg2005symmetries2hdmdifferentvacua}:
{\small
\begin{equation}\label{mostgeneralpotential}
\begin{split}
    V_{\si{2HDM}}&=-\frac{m^2_{11}}{2}\Phi^{\dagger}_{1}\Phi_{1}-\frac{m^2_{22}}{2}\Phi^{\dagger}_{2}\Phi_{2}-\frac{m^2_{12}}{2}\Phi^{\dagger}_{1}\Phi_{2}-\frac{\left(m^2_{12}\right)^{*}}{2}\Phi^{\dagger}_{2}\Phi_{1}+\frac{\lambda_{1}}{2}\left(\Phi^{\dagger}_{1}\Phi_{1}\right)^2+\frac{\lambda_{2}}{2}\left(\Phi^{\dagger}_{2}\Phi_{2}\right)^2\\
    &+\lambda_{3}\left(\Phi^{\dagger}_{1}\Phi_{1}\right)\left(\Phi^{\dagger}_{2}\Phi_{2}\right)+\lambda_{4}\left(\Phi^{\dagger}_{1}\Phi_{2}\right)\left(\Phi^{\dagger}_{2}\Phi_{1}\right)+\frac{\lambda_{5}}{2}\left(\Phi_{1}^{\dagger}\Phi_{2}\right)^2+\frac{\lambda_{5}^{*}}{2}\left(\Phi_{2}^{\dagger}\Phi_{1}\right)^2\\
    &+\lambda_{6}\left(\Phi^{\dagger}_{1}\Phi_{1}\right)\left(\Phi^{\dagger}_{1}\Phi_{2}\right)+\lambda^{*}_{6}\left(\Phi^{\dagger}_{1}\Phi_{1}\right)\left(\Phi^{\dagger}_{2}\Phi_{1}\right)+\lambda_{7}\left(\Phi^{\dagger}_{2}\Phi_{2}\right)\left(\Phi^{\dagger}_{1}\Phi_{2}\right)+\lambda^{*}_{7}\left(\Phi^{\dagger}_{2}\Phi_{2}\right)\left(\Phi^{\dagger}_{2}\Phi_{1}\right)\ .
\end{split}
\end{equation}
}
Note that the normalization selected for the potential in this model is analogous to the one used in the previous example, Eq.~\eqref{DSHM_potential}.

\subsection{\texorpdfstring{$\mathbb{Z}_2$}{Z2}-conserving 2HDM}\label{subsect_softlybrokenZ2}

The most popular variant of the general 2HDM is the one that preserves $\mathbb{Z}_2$ symmetry, that is, invariant under~\cite{Ginzburg_2005_2}
\begin{equation}
    \Phi_1\longleftrightarrow -\Phi_1\ ,\ \Phi_2\longleftrightarrow \Phi_2\ \mbox{or}\ \Phi_1\longleftrightarrow \Phi_1\ ,\ \Phi_2\longleftrightarrow -\Phi_2
\end{equation}
transformations. In order to achieve this, it is necessary to remove $\lambda_{6}$ and $\lambda_{7}$ from the potential, as well as $m^2_{12}$. As in the previous case in Section \ref{section2}, the following change of variables is considered for this much simpler $\mathbb{Z}_2$-conserving 2HDM:
\begin{equation}\label{2HDMvariables}
    x\equiv\Phi^{\dagger}_1\Phi_1\sqrt{\lambda_1}\ \ \ ,\ \ \ y\equiv\Phi^{\dagger}_2\Phi_2\sqrt{\lambda_2}\ .
\end{equation}

However, since there are more possible combinations among fields, we also require
\begin{equation}\label{2HDM_secondvariables}
z\equiv\frac{\left(\lambda_1\lambda_2\right)^{1/4}}{2}\left(\Phi^{\dagger}_2\Phi_1+\Phi^{\dagger}_1\Phi_2\right)\ \ \ ,\ \ \ w\equiv\frac{i\left(\lambda_1\lambda_2\right)^{1/4}}{2}\left(\Phi^{\dagger}_2\Phi_1-\Phi^{\dagger}_1\Phi_2\right)\ .
\end{equation}
Note that $z$ and $w$ can be written in terms of $x$ and $y$. If Equation~\eqref{2HDMvariables} is inverted, 
\begin{equation}\label{Higgs_particles_inverted}
    \Phi_1=\sqrt{x\lambda^{-1/2}_1}n_1\ ,\ \Phi_2=\sqrt{y\lambda^{-1/2}_2}n_2\ ,
\end{equation}
with $n^{\dagger}_1n_1=1$ and $n^{\dagger}_2n_2=1$,
it is possible to ascertain that
\begin{equation}
    z=\sqrt{xy}\Re{\left(n^{\dagger}_1n_2\right)}\ ,\ w=\sqrt{xy}\Im{\left(n^{\dagger}_1n_2\right)}\ .
\end{equation}
Next, we parametrize $n^{\dagger}_1n_2$ as 
\begin{equation}\label{product}
    n^{\dagger}_1n_2=re^{i\phi/2}\ ,\ r\in[0,1]\ ,\ \phi\in[0,4\pi).
\end{equation}
Taking into account all the aforementioned and neglecting quadratic terms, the 2HDM potential can be written as:
{\small
\begin{equation}\label{2HDMpotential_newvariables}
    V^{\mathbb{Z}_2\ (4)}_{\si{2HDM}}=\frac{1}{2}x^2+\frac{1}{2}y^2+xy\left[\rho_3+r^2\left(\rho_4+\|\rho_5\|\cos\left(\alpha_{5}+\phi\right)\right)\right]\ ,
\end{equation}
}
where $\lambda_5=\|\lambda_5\|e^{i\alpha_{5}}$. In virtue of Eq.~\eqref{conditionsinglet} and since Eq.~\eqref{2HDMpotential_newvariables} has the same mathematical structure as Equation~\eqref{DSHMpotential_newparameters}, it is possible to claim that the potential is bounded from below provided
\begin{equation}\label{rho_effective}
\rho^{\si{eff}}_3 (r,\phi)\equiv\rho_3+r^2\left(\rho_4+\|\rho_5\|\cos\left(\alpha_{5}+\phi\right)\right)>-1\ ,
\end{equation}
where,
\begin{equation}
    \rho_{i} = \frac{\lambda_{i}}{\sqrt{\lambda_{1}\lambda_{2}}}\ \ ,\ \ i = 3,4,5.
\end{equation}
Even though the RSE and the $\mathbb{Z}_2$-conserving 2HDM might seem to differ when written in terms of the fields, given the correct changes of variables ---Equations~\eqref{first_changeofvariable} and \eqref{2HDMvariables}--- both potentials reduce to a similar structure. Nonetheless, Equation~\eqref{2HDMpotential_newvariables} depends on how both Higgs doublets are coupled; therefore, these two particles will eventually end up in the lowest energy state ---which corresponds, in this case, to the lowest $\rho^{\si{eff}}_3$. Hence, we urge to find $\rho^{\si{eff}}_{3,\si{min}}$ and impose $\rho^{\si{eff}}_{3,\si{min}}>-1$, instead of Equation~\eqref{rho_effective}, to ensure that it is impossible for both Higgs doublets to temporarily evolve into an unbounded lower energy state ($\rho^{\si{eff}}_3<-1$ for certain $r^2$ and $\phi$). In order to achieve this, consider the rectangle $[0,1]\times[-1,1]$ where $\rho^{\si{eff}}_3(r^2,\cos(\alpha_{5}+\phi))$ is defined, and let $A$, $B$, $C$ and $D$ be its borders:

 $$A=\{(r^2,\cos(\alpha_{5}+\phi)):\ r^2=0\ ,\ \cos(\alpha_{5}+\phi)\in [-1,1]\}$$
 
 $$B=\{(r^2,\cos(\alpha_{5}+\phi)):\ \cos(\alpha_{5}+\phi)=1\ ,\ r^2\in(0,1)\}$$
 
 $$C=\{(r^2,\cos(\alpha_{5}+\phi)):\ r^2=1\ ,\ \cos(\alpha_{5}+\phi)\in [-1,1]\}$$
 
 $$D=\{(r^2,\cos(\alpha_{5}+\phi)):\ \cos(\alpha_{5}+\phi)=-1\ ,\ r^2\in(0,1)\}$$

It is straightforward to demonstrate that there are no absolute maxima or minima inside the above-mentioned rectangle. Consequently, minima are found in the borders. In fact, through a naive reasoning, it is demonstrable that $\rho^{\si{eff}}_{3,\si{min}}$ can be found in
\begin{equation}
\begin{split}
&\mbox{(Case 1)}\ \ A,\ \ \mbox{if}\ \ \rho_4\geq\|\rho_5\|.\\
\\
&\mbox{(Case 2)}\ \ A,\ \ D\ \ \mbox{and}\ \ (r^2,\cos(\alpha_{5}+\phi))=(1,-1)\in C,\ \ \mbox{if}\ \ \rho_4=\|\rho_5\|.\\
\\
&\mbox{(Case 3)}\ \ C,\ \ \mbox{if}\ \ \rho_4<\|\rho_5\|=0.\\
\\
&\mbox{(Case 4)}\ \ (r^2,\cos(\alpha_{5}+\phi))=(1,-1)\in C,\ \ \mbox{otherwise}.\\
\end{split}
\end{equation}
Regardless of the case, the value of $\rho^{\si{eff}}_{3,\si{min}}$ is always given by
\begin{equation}
\label{rho_3_minimum}    \rho^{\si{eff}}_{3,\si{min}}=\rho_3+\min{\left(\rho_4-\|\rho_5\|,0\right)}\ .
\end{equation}
As mentioned before, a more appropriate condition to the parameters of the potential is
\begin{equation}\label{real_cond}
\rho^{\si{eff}}_{3,\si{min}}>-1\ \left(\mbox{also}\ \lambda_{1,2}>0,\ \mbox{see Section\ \ref{section2}}\right)\ ,
\end{equation}
instead of Equation~\eqref{rho_effective}. Given the restriction of Eq.~\eqref{real_cond}, one can deduce the following:
\begin{equation}\label{cond_over_NO_effective_rho3}
\left\{
\begin{array}{rcl}
    &\rho_3 &> -1,\ \mbox{always} \\
    &\rho_3 &> -1 -\rho_4,\ \mbox{in the projection onto }\rho_3\ \mbox{and }\rho_4\ .
\end{array}
\right.
\end{equation}

\begin{figure}[p]
    \centering
    \includegraphics[scale=0.4]{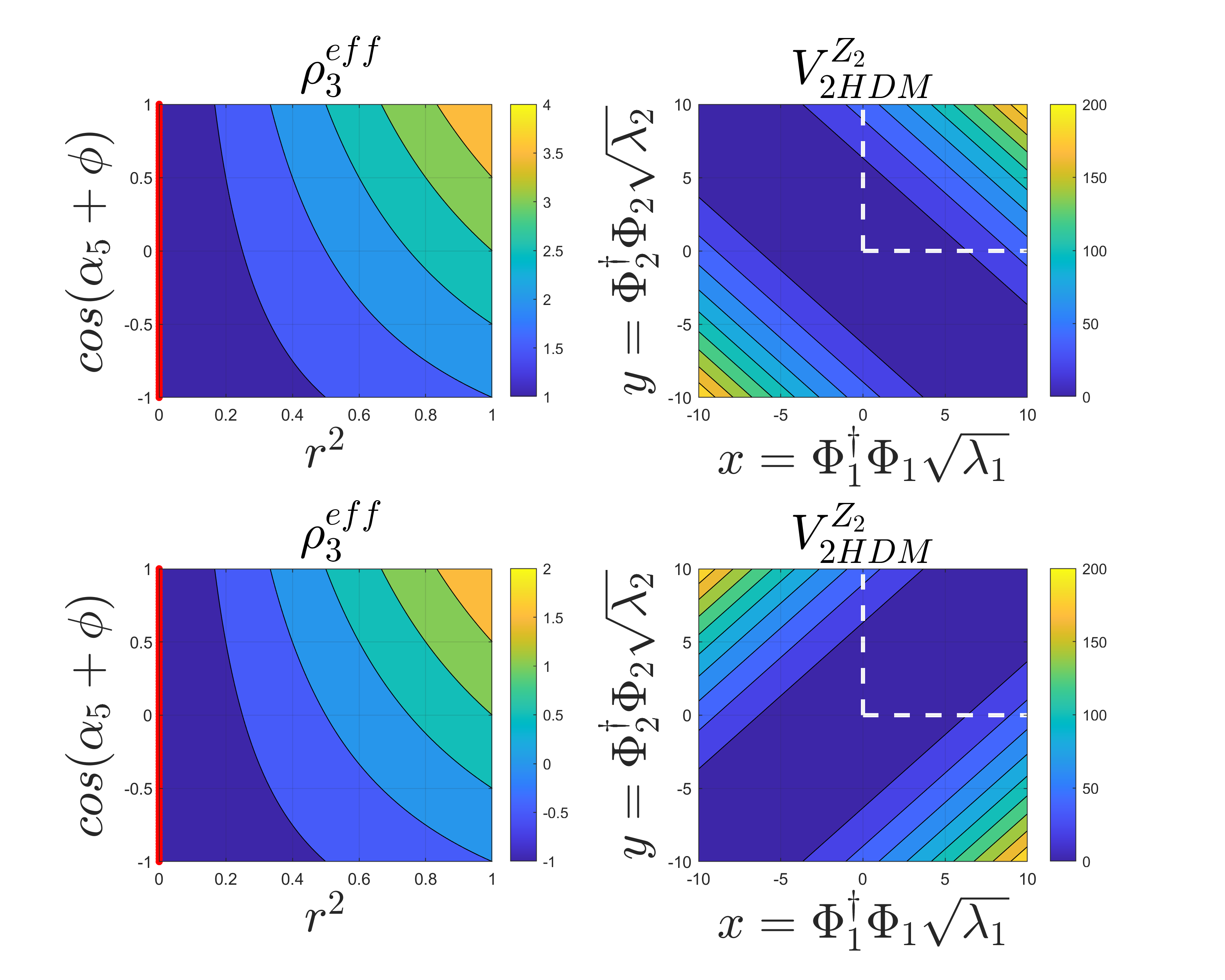} 
    \caption{\textit{Case 1}, \textit{Left}: $\rho^{\si{eff}}_3$ in the rectangle $[0,1]\times[-1,1]$. Absolute minima are highlighted in red. All those points generate the potential $V^{\mathbb{Z}_2}_{2\si{HDM}}(\rho^{\si{eff}}_{3,\si{min}})$ ---see the corresponding figure on the right. The potential in the upper row is valid because $\rho^{\si{eff}}_{3,\si{min}}>-1$, while the potential in the lower row is not physically possible. \textit{Upper row}: $\left(\rho_3,\rho_4,\|\rho_5\|,\alpha_{5}\right)=\left(1,2,1,0\right)$. \textit{Lower row}: $\left(\rho_3,\rho_4,\|\rho_5\|,\alpha_{5}\right)=\left(-1,2,1,0\right)$.}
    \label{Case_1_2HDM}
\end{figure}

\begin{figure}[p]
    \centering
    \includegraphics[scale=0.4]{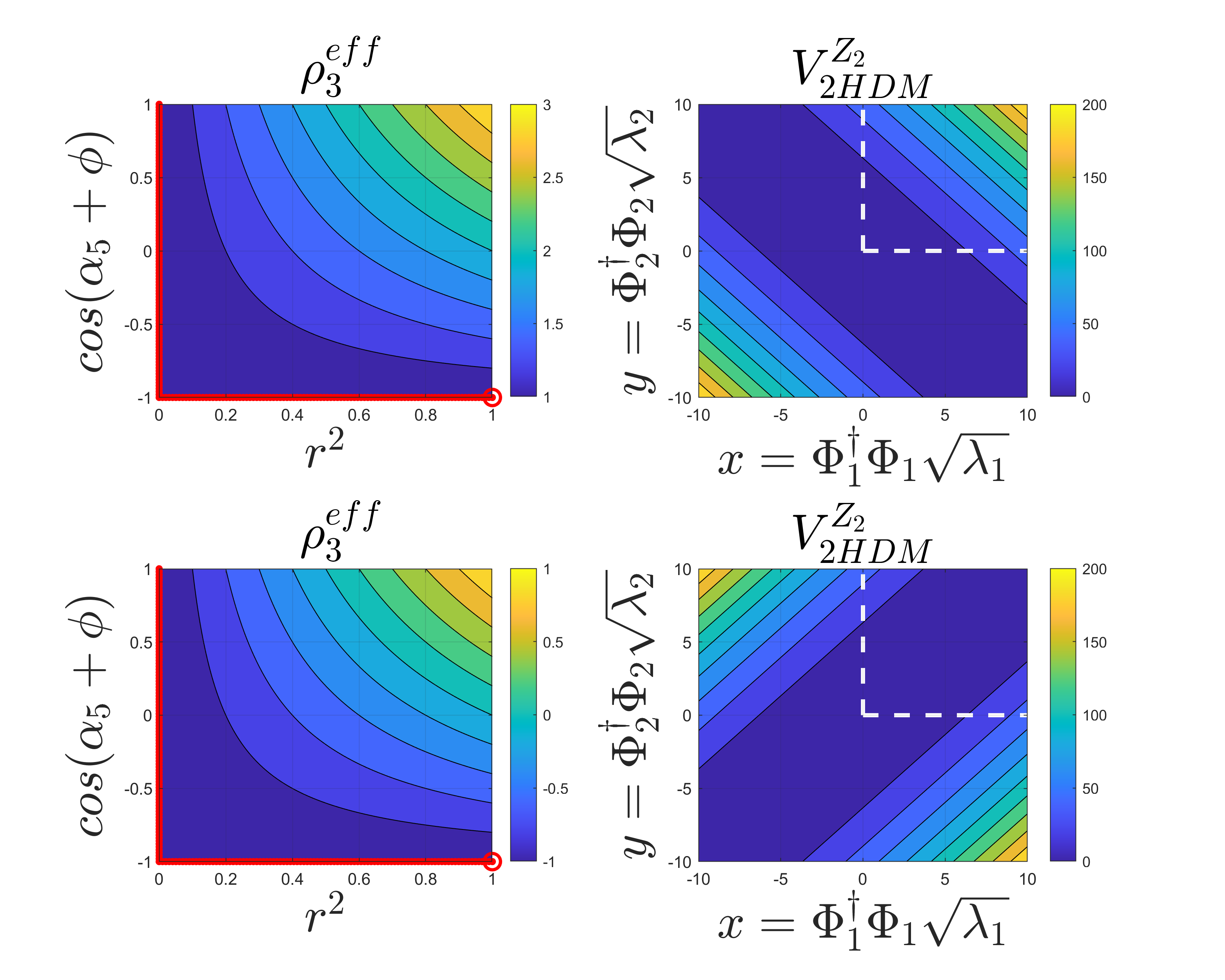} 
    \caption{\textit{Case 2}, \textit{Left}: $\rho^{\si{eff}}_3$ in the rectangle $[0,1]\times[-1,1]$. Absolute minima are highlighted in red. All those points generate the potential $V^{\mathbb{Z}_2}_{2\si{HDM}}(\rho^{\si{eff}}_{3,\si{min}})$ ---see the corresponding figure on the right. The potential in the upper row is valid because $\rho^{\si{eff}}_{3,\si{min}}>-1$, while the potential in the lower row is not physically possible. \textit{Upper row}: $\left(\rho_3,\rho_4,\|\rho_5\|,\alpha_{5}\right)=\left(1,1,1,0\right)$. \textit{Lower row}: $\left(\rho_3,\rho_4,\|\rho_5\|,\alpha_{5}\right)=\left(-1,1,1,0\right)$.}
    \label{Case_2_2HDM}
\end{figure}

\begin{figure}[!t]
    \centering
    \includegraphics[scale=0.4]{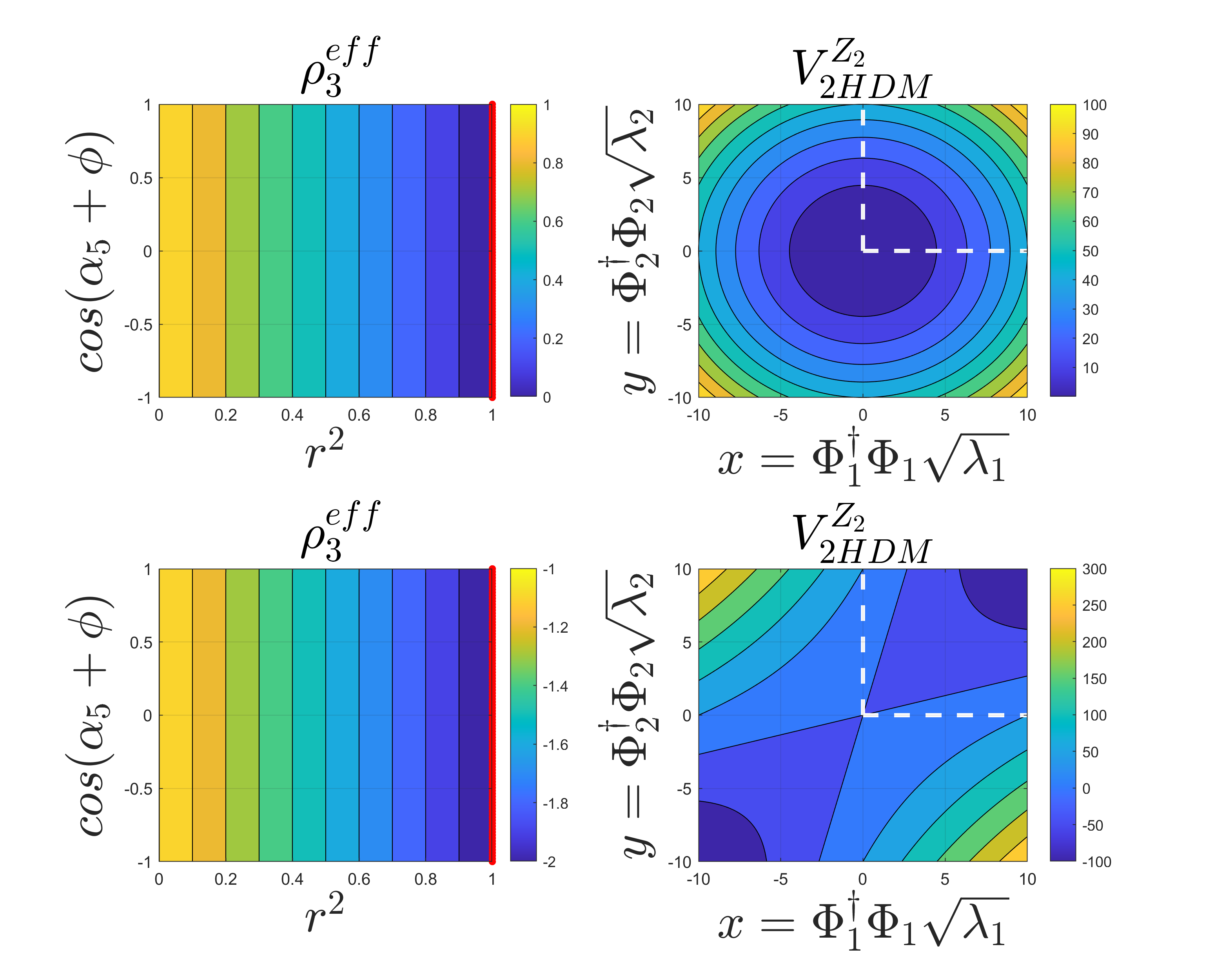} 
    \caption{\textit{Case 3}, \textit{Left}: $\rho^{\si{eff}}_3$ in the rectangle $[0,1]\times[-1,1]$. Absolute minima are highlighted in red. All those points generate the potential $V^{\mathbb{Z}_2}_{2\si{HDM}}(\rho^{\si{eff}}_{3,\si{min}})$ ---see the corresponding figure on the right. The potential in the upper row is valid because $\rho^{\si{eff}}_{3,\si{min}}>-1$, while the potential in the lower row is not physically possible. \textit{Upper row}: $\left(\rho_3,\rho_4,\|\rho_5\|,\alpha_{5}\right)=\left(1,-1,0,0\right)$. \textit{Lower row}: $\left(\rho_3,\rho_4,\|\rho_5\|,\alpha_{5}\right)=\left(-1,-1,0,0\right)$.}
    \label{Case_3_2HDM}
\end{figure}

\begin{figure}[!t]
    \centering
    \includegraphics[scale=0.4]{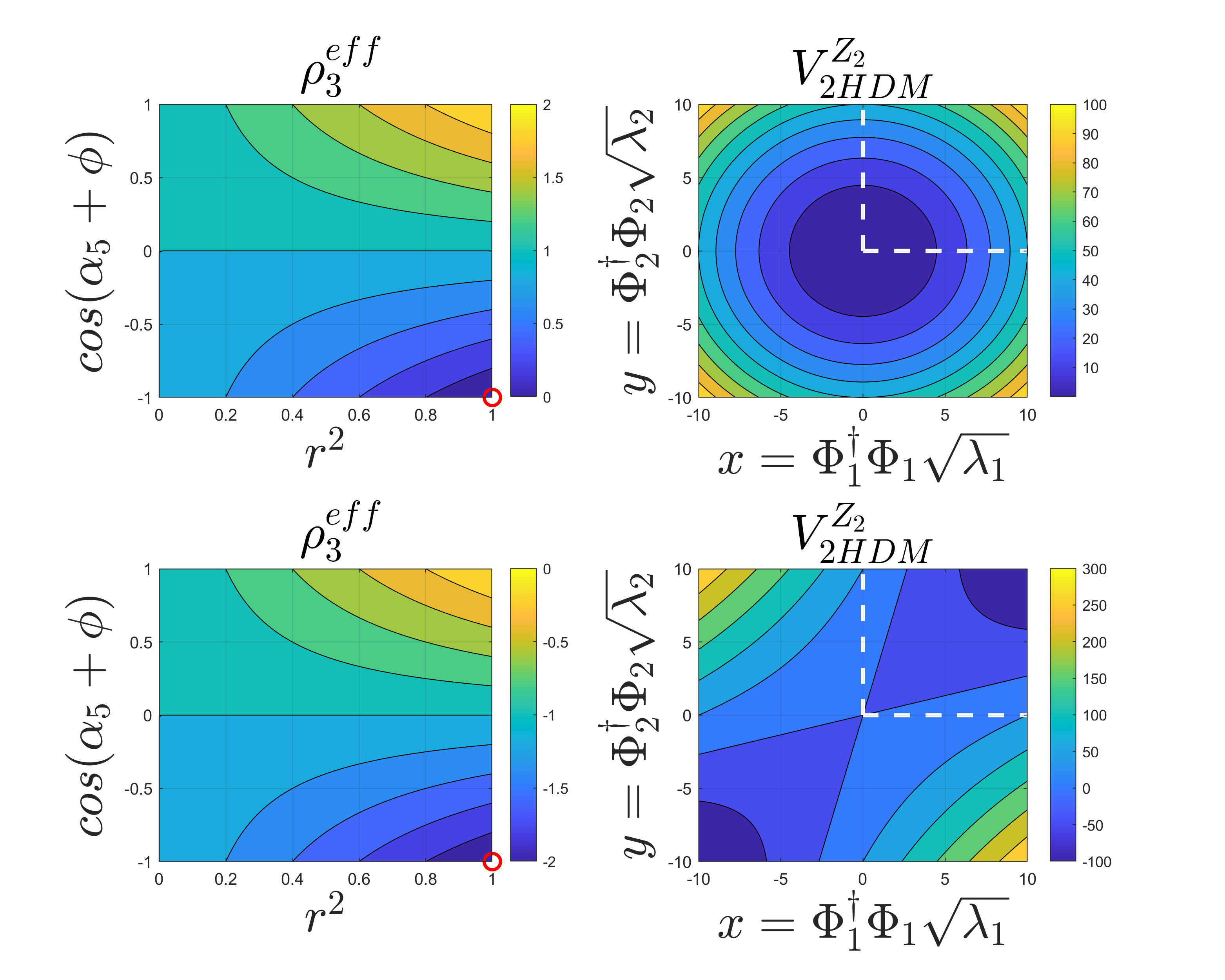} 
    \caption{\textit{Case 4}, \textit{Left}: $\rho^{\si{eff}}_3$ in the rectangle $[0,1]\times[-1,1]$. The absolute minimum is highlighted in red. That point generates the potential $V^{\mathbb{Z}_2}_{2\si{HDM}}(\rho^{\si{eff}}_{3,\si{min}})$ ---see the corresponding figure on the right. The potential in the upper row is valid because $\rho^{\si{eff}}_{3,\si{min}}>-1$, while the potential in the lower row is not physically possible. \textit{Upper row}: $\left(\rho_3,\rho_4,\|\rho_5\|,\alpha_{5}\right)=\left(1,0,1,0\right)$. \textit{Lower row}: $\left(\rho_3,\rho_4,\|\rho_5\|,\alpha_{5}\right)=\left(-1,0,1,0\right)$.}
    \label{Case_4_2HDM}
\end{figure}

These conditions can also be found in Refs.~\cite{Deshpande:1977rw,Ginzburg_2005,Maniatis:2006fs,Kannike:2012pe,Kannike_2016,Klimenko:1984qx,Nie:1998yn,Kanemura:1999xf}. Note that, even though there may exist other lower-order terms which violate the $\mathbb{Z}_2$ symmetry, this result remains valid since quartic terms have the structure outlined in Equation~\eqref{structure}. Therefore, all 2HDM with a soft $\mathbb{Z}_2$ symmetry breaking must comply with Equation~\eqref{real_cond}. Similarly to Figure~\ref{DSHM}, one can observe in Figures~\ref{Case_1_2HDM}--\ref{Case_4_2HDM} that each case must obey Equation~\eqref{real_cond}, otherwise, if $\rho^{\si{eff}}_{3,\si{min}}\leq-1$, the potential decreases in the first quadrant and, consequently, cannot be physically valid.

\begin{figure}[!t]
    \centering
    \includegraphics[scale=0.35]{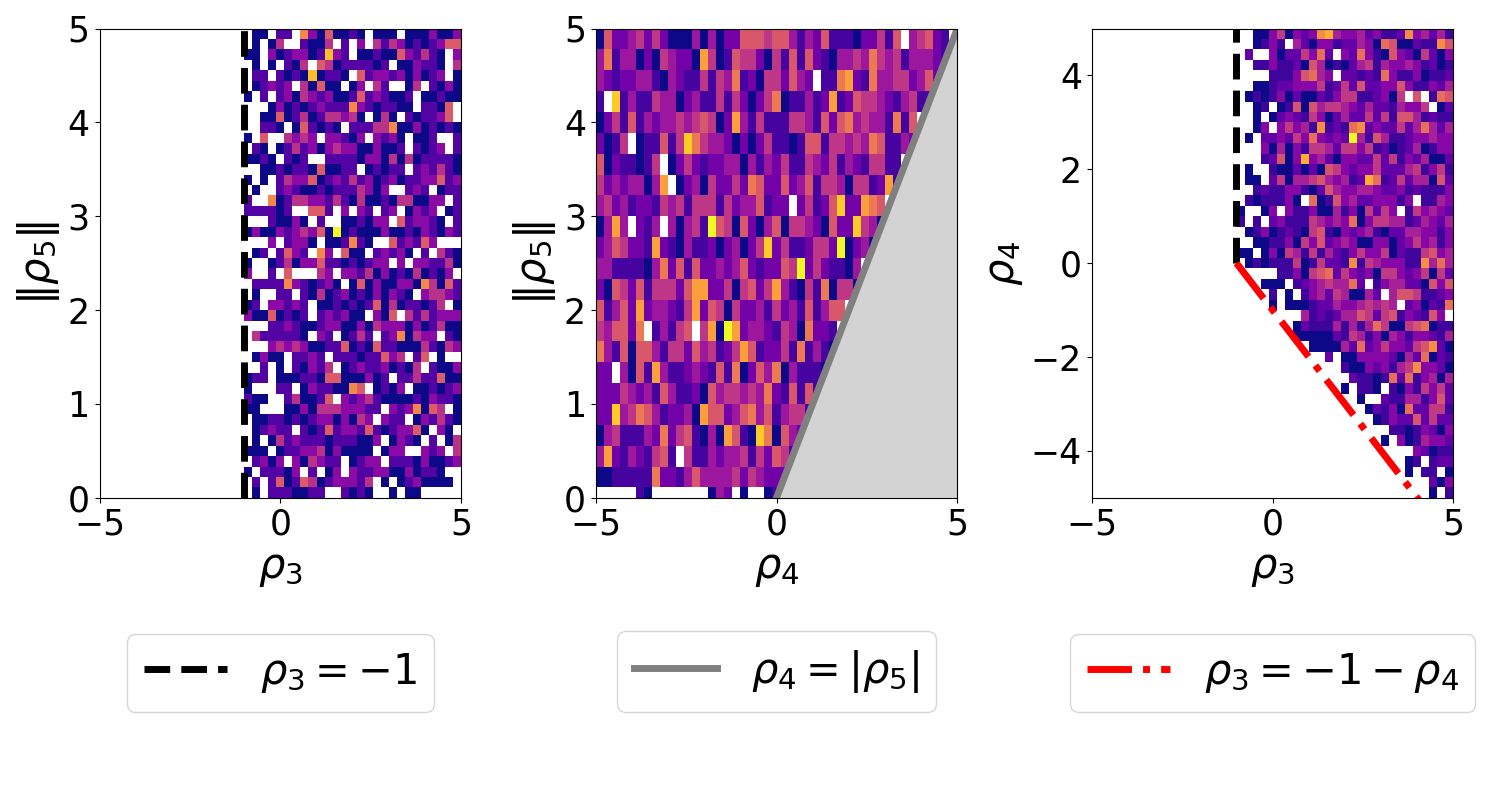} 
    \caption{Theoretical restrictions of Eq.~\eqref{cond_over_NO_effective_rho3} verified computationally. Note that the dashed black line corresponds to $\rho_{3}=-1$, while the dash-dotted red line, to the restriction $\rho_{3}=-1-\rho_{4}$. Equation~\eqref{vacionnocargado} has also been included, see gray line. If the algorithm applied only imposes the positivity of the potential, there will be ``valid'' points in the $\rho_{4}>\|\rho_{5}\|$ region ---manually filled with light gray color.}
    \label{Z2_projection_rho_345}
\end{figure}

Moreover, the restrictions in Equation~\eqref{cond_over_NO_effective_rho3} can be corroborated numerically with our public Rust code~\cite{javi}. This is shown in Figure~\ref{Z2_projection_rho_345}, where there exists a complete agreement between the theoretical and the computational analyses. To the best of our knowledge, no similar numerical approach to this problem has been provided in the literature with the level of detail in this article, which also studies the allowed Lagrangian parameters in the $\mathbb{Z}_2$-broken scenario. 

At the same time, additional constraints besides the positivity of the potential can be considered. For example, a reasonable restriction to contemplate is to impose that the photon is massless. This can be achieved through the kinetic energy term in this $\mathbb{Z}_2$-conserving model:
\begin{equation}\label{kinetic_term}
    T^{\mathbb{Z}_2}_{2\si{HDM}}=\left(D_{\mu}\Phi_1\right)^{\dagger}\left(D^{\mu}\Phi_1\right)+\left(D_{\mu}\Phi_2\right)^{\dagger}\left(D^{\mu}\Phi_2\right)\ ,
\end{equation}
with
\begin{equation}
    D_{\mu}=\partial_{\mu}+\frac{ig}{2}W^{a}_{\mu}\tau^{a}+\frac{ig'}{2}B_{\mu}\ ,
\end{equation}
where $\tau^{a}$ are Pauli matrices. The quadratic terms of Equation~\eqref{kinetic_term} are considered as mass terms. This part of the kinetic term can be rewritten as:
\begin{equation}
    T^{\mathbb{Z}_2}_{2\si{HDM},\si{mass}}=\frac{1}{2}\begin{pmatrix}
        W^{a}_{\mu} & B_{\mu}
    \end{pmatrix}\mathcal{M}^{ab}\begin{pmatrix}
        W^{b\mu}\\
        B^{\mu}
    \end{pmatrix}\ .
\end{equation}

In order to have a massless photon, one must assure that the quadratic form $\mathcal{M}^{ab}$ has one eigenvalue that is equal to zero. With these guidelines, except for inert vacua where one of the doublets
has a vanishing vev, an additional restriction to the parameter space can be found \cite{Diaz-Cruz:1992fcx}:
\begin{equation}
    \|\lambda_{5}\|\geq\lambda_{4}\ ,
\end{equation}
which can be translated as
\begin{equation}\label{vacionnocargado}
    \|\rho_{5}\|\geq\rho_{4}\ .  
\end{equation}
This result has also been considered in Fig.~\ref{Z2_projection_rho_345}, which provides an additional condition that narrows down the parameter valid region.  Nevertheless, we want to point out that the Rust code we provide only analyzes positivity restrictions emerging from the potential. Additionally, the output parameters from our code that violated Eq.~\eqref{vacionnocargado} were manually removed (gray region). However, this additional restriction is not required for all parameters providing a massless photon. In particular, inert vacua do not need this additional condition. Consequently, the gray region in Fig.~\ref{Z2_projection_rho_345} would be allowed by the previous BFB conditions. Finally, note that this restriction ---Eq.~\eqref{vacionnocargado}--- might not be valid when considering the $\mathbb{Z}_2$-violating 2HDM. Nevertheless, this is beyond the scope of this work and this vacuum constraint will not be further discussed.
 
\subsection{\texorpdfstring{$\mathbb{Z}_2$}{Z2}-violating 2HDM}\label{section4}

In this section, we aim to scrutinize the most general 2HDM \footnote{The interest in studying the $\mathbb{Z}_2$-violating 2HDM potentials mainly lies in the fact that they allow the existence of flavor-changing neutral currents (FCNC) \cite{Branco_2012,Davidson:2005cw,Hou:1991un,Atwood_1997}. On top of that, the parameters that arise when the $\mathbb{Z}_2$ symmetry is broken are complex. Hence, this might be an additional mechanism for CP violation \cite{Branco_2012,ginzburg2003treelevelunitarityconstraints2hdm,ginzburg2005symmetries2hdmdifferentvacua,Ginzburg_2005,Ginzburg_2005_2,Boto_2020,Barger_2009}. However, these topics exceed the scope of this work.} ---the one without $\mathbb{Z}_2$ symmetry, herein indicated as $\backslash\hspace*{-0.23cm}\mathbb{Z}_2$--- whose potential has been written in Equation~\eqref{mostgeneralpotential}. 
The forthcoming step consists in expressing this potential using $x$ and $y$ variables of Eq.~\eqref{2HDMvariables}. In light of the following definitions,
\begin{equation}
\lambda_{j}\equiv\|\lambda_j\|e^{i\alpha_{j}}\ \ \ ,\ \ \ \|\rho_{j}\|\equiv 2\|\lambda_{j}\|\lambda^{-1/4}_{1}\lambda^{-1/4}_{2}\lambda^{-1/2}_{j-5}\ \ \ ,\ \ \ \mbox{with}\ \ \ j=6,7\ .
\end{equation}
and the aforementioned definitions of $x$ and $y$, it is possible to write the quartic terms of the most general 2HDM potential as
\begin{equation}\label{simplerV2_Z2_less_pot}
    V^{\backslash\hspace*{-0.15cm}\mathbb{Z}_2\ (4)}_{2\si{HDM}}=\frac{1}{2}x^2+\frac{1}{2}y^2+\rho^{\si{eff}}_{3}xy+\rho^{\si{eff}}_{6}x\sqrt{xy}+\rho^{\si{eff}}_{7}y\sqrt{xy}\ ,
\end{equation}
where
\begin{equation}
\rho^{\si{eff}}_6 \equiv\|\rho_{6}\|r\cos{\left(\alpha_{6}+\frac{\phi}{2}\right)}\ \ \ \mbox{and}\ \ \ \rho^{\si{eff}}_7 \equiv\|\rho_{7}\|r\cos{\left(\alpha_{7}+\frac{\phi}{2}\right)}.
\end{equation}
Note that  $\phi$ ---which provides the relative phase between $\Phi_1$ and $\Phi_2$--- only enters the potential through the combinations $\phi+\alpha_5$, $\phi+ 2\alpha_6$ and $\phi+2\alpha_7$ in $\rho_3^{\rm eff}$, $\rho_6^{\rm eff}$ and $\rho_7^{\rm eff}$, respectively.

Next, we introduce our polar coordinates ---see Equation~\eqref{ourpolarcoordinates}. Due to this, one can write Equation~\eqref{simplerV2_Z2_less_pot} as follows:
\begin{equation}\label{2hdm_polar}
V^{\backslash\hspace*{-0.15cm}\mathbb{Z}_2\ (4)}_{2\si{HDM}}=\frac{t^4}{4}\left[2+\left(\rho^{\si{eff}}_{3}-1\right)\sin^{2}{\left(2\chi\right)}+\left(u+v\cos{\left(2\chi\right)}\right)\sin{\left(2\chi\right)}\right] \equiv \frac{t^4}{4}f(\chi) >0\ ,
\end{equation}
where some angle identities have been used, as well as
\begin{equation}\label{rho6and7def}
u\equiv\rho^{\si{eff}}_{6}+\rho^{\si{eff}}_{7}\ \ \ \mbox{and}\ \ \ v\equiv\rho^{\si{eff}}_{6}-\rho^{\si{eff}}_{7}\ .
\end{equation}

It is interesting to note that if $\rho^{\si{eff}}_{6}=\rho^{\si{eff}}_{7}=0$, the results of Equation~\eqref{rho_effective} are recovered. Next, we study three cases within the $\mathbb{Z}_2$-violating scenario, depending on the values of $\rho^{\si{eff}}_{6}$ and $\rho^{\si{eff}}_{7}$ that we have called: symmetric ($v=0$), antisymmetric ($u=0$) and the most general potential.

\subsubsection{Symmetric \texorpdfstring{$\mathbb{Z}_2$}{Z2}-violating potential}

\begin{figure}[!t]
    \centering
    \includegraphics[scale=0.35]{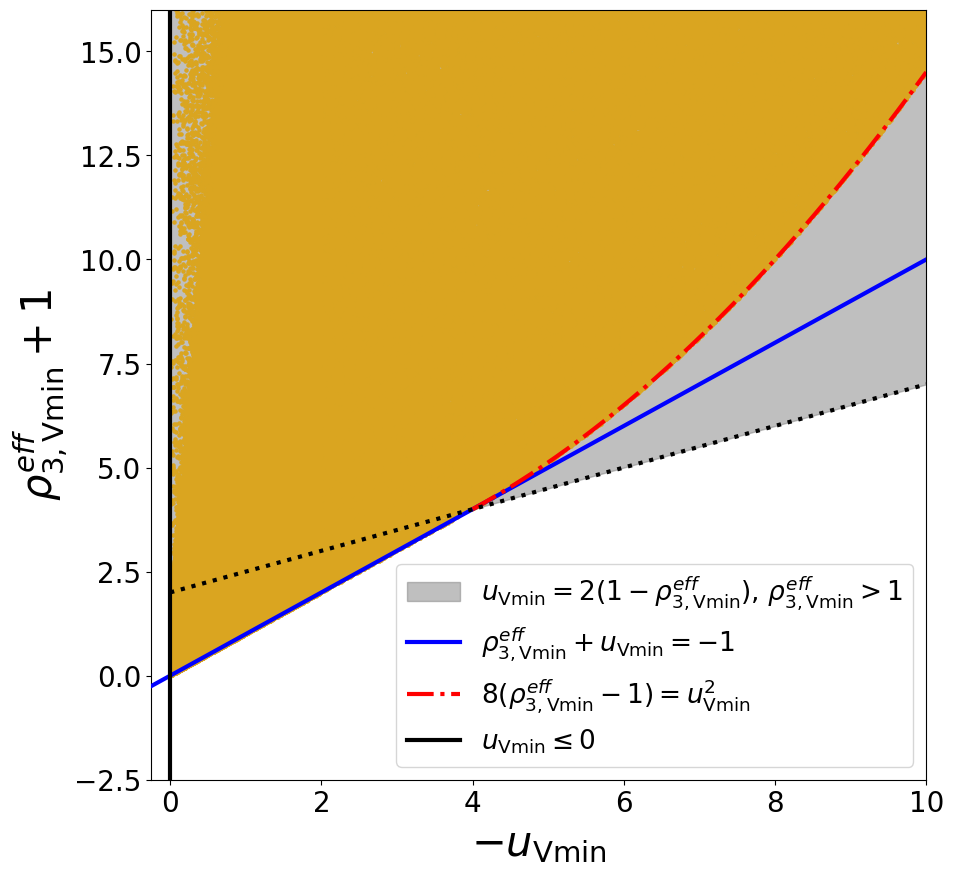} 
    \caption{The restrictions for this symmetric 2HDM subcase presented in Equation~\eqref{minimo_hs} have been corroborated computationally \cite{javi} and plotted ---yellow dots. This plot shows that the pair of restrictions in Equation~\eqref{minimo_hs} can be rewritten as only one condition: a straight line connected to a parabola. Note that the parabola can only be considered within the region with a gray background.}
    \label{sym_restiction}
\end{figure}

One subcase is the symmetric $\mathbb{Z}_2$-violating potential, that is to say, the scenario where $\rho^{\si{eff}}_{6}=\rho^{\si{eff}}_{7}\neq 0$ ($v=0$) in Equation~\eqref{2hdm_polar}. Hence, the angular part of the potential is
\begin{equation}\label{sympot_conclusions}
    f(s) = 2+\left(\rho^{\si{eff}}_{3}-1\right)s^2+us>0\ ,
\end{equation}
where $s\equiv\sin{(2\chi)}\in (0,1]$.

We aim to find the infimum for this parabola $f(s)$ in $(0,1]$ and impose that it be greater than zero. Since $\lim\limits_{s\to 0} f(s)=2>0$ always, we only have to impose $f(1)>0$ or that the value of the function in the extremum be greater than zero when necessary. This constrains the parameter space as follows:
\begin{equation}\label{minimo_hs}
    \left\{\begin{array}{rlc}
    \vspace{5mm}
    & \rho^{\si{eff}}_{3,\si{V}_{\si{min}}}+u_{\si{V}_{\si{min}}}\ >\ -1\ ,\ \mbox{always}\\
    \vspace{1mm}
    & 8\left(\rho^{\si{eff}}_{3,\si{V}_{\si{min}}}-1\right)\ >\  u^2_{\si{V}_{\si{min}}},\,\, \mbox{if}\ u_{\si{V}_{\si{min}}}\in\left[ 2\left(1-\rho^{\si{eff}}_{3,\si{V}_{\si{min}}}\right),0\right]\ \mbox{and}\ \rho^{\si{eff}}_{3,\si{V}_{\si{min}}}>1
    \end{array}\right.
\end{equation}
Besides these constraints of Eq.~\eqref{minimo_hs}, there is an additional theoretical restriction that must be applied:
\begin{equation}\label{umenorzero}
    u_{\si{V}_{\si{min}}}\leq 0.
\end{equation}
To understand the reason behind this new condition, let us suppose that $u>0$ and all the parameters of the potential remain unchanged. Given the definitions of $\rho^{\si{eff}}_{3}$ and $u$ ---Equations \eqref{rho_effective} and \eqref{rho6and7def}, in this scenario it is always possible to apply the shift $\phi\longrightarrow\phi + 2\pi$, maintaining $r$ and $\chi$ constant, and obtain the same value of $\rho^{\si{eff}}_{3}$ but a negative value of $u$, which gives a lower value of the potential. This makes the latter a better candidate for the value of $u$ that minimizes the potential: $u_{\si{V}_{\si{min}}}$. In fact, the value of $u$ after the phase shift is $-u$.

Once all the theoretical restrictions have been presented, it is possible to corroborate them computationally \cite{javi}, as it has been done in Figure~\ref{sym_restiction}. Note that $\rho^{\si{eff}}_{3}$ and $u$ must be evaluated at the values of $r$ and $\phi$ that minimize the potential from Eq.~\eqref{simplerV2_Z2_less_pot}. In Figure~\ref{sym_restiction}, one can observe an agreement between the numerical and the theoretical analyses as all points lie above the blue line, which is the first constraint of Equation~\eqref{minimo_hs}. Similarly, all points are bounded on the left by Equation~\eqref{umenorzero}. However, the second constraint of Equation~\eqref{minimo_hs} is only applicable under certain conditions. Those two conditions, see Equation~\eqref{minimo_hs}, can be visualized as the region highlighted with a gray background. As a point of interest, the application of the second constraint of Equation~\eqref{minimo_hs} starts in the intersection of the first and the second constraint, therefore, Equation~\eqref{minimo_hs} can be combined into a single condition consisting of a straight line joined to a parabola. It is worth stressing that similar results do not seem to appear in the literature, up to our knowledge.   

Unlike in the $\mathbb{Z}_2$-conserving 2HDM, in this model with additional degrees of freedom, the minimum of the potential might not be reached when considering the minimum values of $\rho^{\si{eff}}_{i}\ i=3,6,7$. This is the reason behind the new notation with the subscript $V_{\si{min}}$. 

\subsubsection{Antisymmetric \texorpdfstring{$\mathbb{Z}_2$}{Z2}-violating potential}

The second subcase is the antisymmetric $\mathbb{Z}_2$-violating potential, where $\rho^{\si{eff}}_{6}=-\rho^{\si{eff}}_{7}\neq 0$ ($u=0$). In this case, the angular part of the potential can be written as \footnote{Note that some half-angle and double-angle relations have been used in Equation~\eqref{2hdm_polar} to derive this result of Eq.~\eqref{antisym_pot}, in particular, $2\sin^2{\left(2\chi\right)}=1-\cos{\left(4\chi\right)}$. Moreover, it is crucial to highlight the fact that even though $\sin{\chi}$ and $\cos{\chi}$ must be greater than zero, due to the choice of coordinates that has been considered, the values of $\sin{\left(4\chi\right)}$ and $\cos{\left(4\chi\right)}$ are not restricted.}:
\begin{equation}\label{antisym_pot}
    2f(\chi) = \rho^{\si{eff}}_{3}+3-\left(\rho^{\si{eff}}_{3}-1\right)\cos{\left(4\chi\right)}+v\sin{\left(4\chi\right)}>0\ .
\end{equation}

Now, we urge to find the minima of $2f(\chi)$. Extrema are found to be $\chi_{\si{ext}}$, that obey:
\begin{equation}\label{condition_extrema_antisym_pot}
    \tan{\left(4\chi_{\si{ext}}\right)}=\frac{v}{1-\rho^{\si{eff}}_{3}}\equiv\zeta\ .
\end{equation}

\begin{figure}[!t]
    \centering
    \includegraphics[scale=0.35]{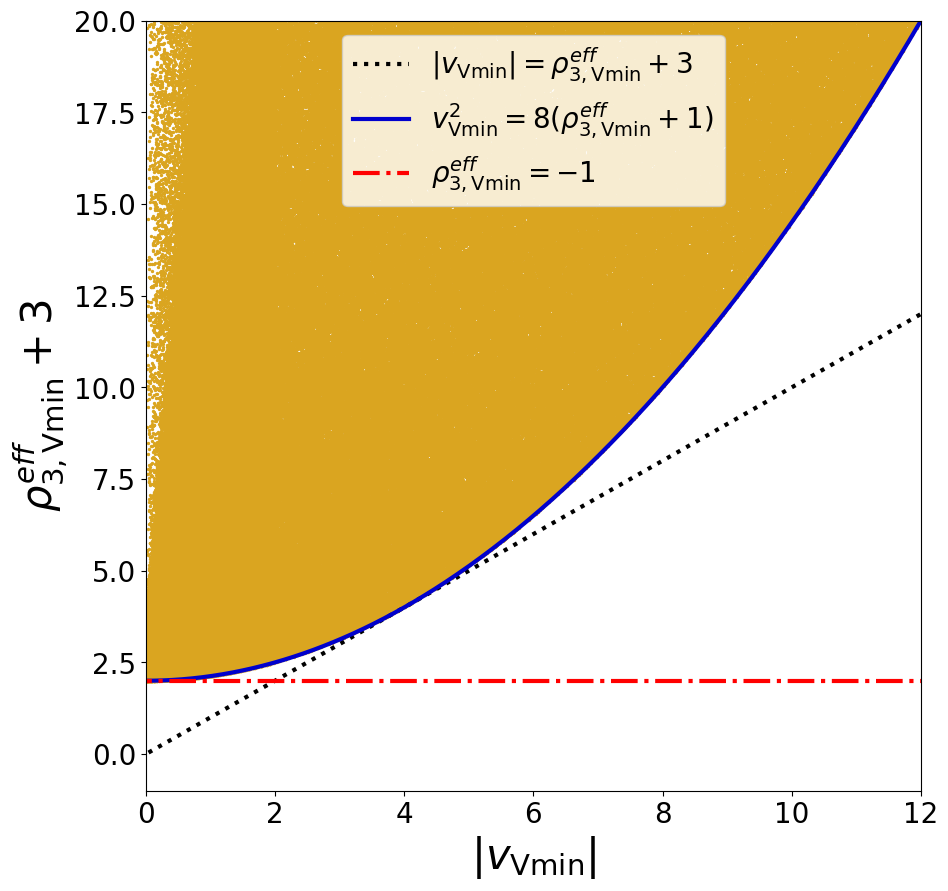} 
    \caption{The restrictions for this antisymmetric 2HDM subcase presented in Equations \eqref{antisym_border_restriction} and \eqref{ann} have been corroborated computationally \cite{javi} and plotted ---yellow dots. This plot shows that the pair of restrictions in Equation~\eqref{antisym_border_restriction} are tangent to the condition in Equation~\eqref{ann}.}
    \label{antisym_restriction}
\end{figure}

During the following derivations, these two identities ---valid in the first quadrant--- will be used:
\begin{equation}\label{identities}
\sin{\left(\arctan{x}\right)}=\frac{x}{\sqrt{1+x^2}}\ \ \ \mbox{and}\ \ \ \cos{\left(\arctan{x}\right)}=\frac{1}{\sqrt{1+x^2}}.
\end{equation}
Note that in each quadrant it is possible to find an expression similar to Equation~\eqref{identities}, provided the signs of sine and cosine are taken into account. Furthermore, the fact that Equation~\eqref{condition_extrema_antisym_pot} has two types of solutions ($\arctan{\zeta}$ and $\arctan{\zeta}+\pi$) must also be considered. However, this phase change that appears when studying both types of solutions affects the sign of the trigonometric functions in each quadrant simultaneously. Hence, the idea is to first generalize Equation~\eqref{identities} for every quadrant, then adjust the signs in regards of the solution of Equation~\eqref{condition_extrema_antisym_pot} that is being analyzed in that moment and finally impose $f(\chi_{\si{min}})>0$. This leaves us with:
\begin{equation}\label{fmin}
    2f(\chi_{\si{min}})=\rho^{\si{eff}}_{3}+3+\frac{1}{\sqrt{1+\zeta^2}}\left(1-\rho^{\si{eff}}_{3}\right)\cdot\left(\sgn{\left(\cos{\left(4\chi_{\si{min}}\right)}\right)}+\sgn{\left(\sin{\left(4\chi_{\si{min}}\right)}\right)}\zeta^2\right)>0\ .
\end{equation}
The interval $(0,2\pi)$ has been divided in four smaller intervals ---one for each quadrant--- and the parameters of the potential must obey Equation~\eqref{fmin} in order to be valid. However, the minimum values are not always found inside those intervals, but sometimes in the borders. Therefore, in the axes separating the quadrants, we have to impose
\begin{equation}\label{restriction_borders}
\begin{split}
    &\ \ \rho^{\si{eff}}_{3}>-1\\
    &\left(\rho^{\si{eff}}_{3}-1\right)\left(1-\zeta\right)>-4\\
    &\left(\rho^{\si{eff}}_{3}-1\right)\left(1+\zeta\right)>-4
\end{split}
\end{equation}
when we move from one quadrant to the adjacent one (in $\chi=0,\ \pi/8,\ \pi/4,\ 3\pi/8\ \mbox{and}\ \pi/2$) so as to guarantee that in the borders the potential is still valid. After simple manipulations, Equation~\eqref{restriction_borders} becomes 
\begin{equation}\label{antisym_border_restriction}
    \begin{split}
        &\ \ \rho^{\si{eff}}_{3,\si{V}_{\si{min}}}>-1\\
        &|v_{\si{V}_{\si{min}}}|<3+\rho^{\si{eff}}_{3,\si{V}_{\si{min}}}\ .
    \end{split}
\end{equation}
Now, if the same analysis is performed inside all four quadrants (not the borders), one can ascertain that the worst-case scenario for Equation~\eqref{fmin} can be written as:
\begin{equation}\label{absqrt}    
    2f(\chi_{\si{min}})=\left. 4+\left(\rho^{\si{eff}}_{3}-1\right)-\left|1-\rho^{\si{eff}}_{3}\right|\sqrt{1+\zeta^2}\right|_{\si{V}_{\si{min}}}>0\ .
\end{equation}

These final equations constitute the constraints governing this model. Going one step further, in light of Equation~\eqref{condition_extrema_antisym_pot}, the Equation~\eqref{absqrt} can be rewritten as
\begin{equation}\label{ann}
    v_{\si{V}_{\si{min}}}^2<8\left(\rho^{\si{eff}}_{3,\si{V}_{\si{min}}}+1\right)\ .
\end{equation}
Note that $\rho^{\si{eff}}_{3}$ and $v$ must be evaluated at the values of $r$, $\phi$ and $\chi$ that minimize the energy of Eq.~\eqref{simplerV2_Z2_less_pot}. In Fig.~\ref{antisym_restriction}, generated by our publicly available Rust code~\cite{javi}, it is also possible to observe an agreement between the expressions derived and the computational analysis. Additionally, in view of this graphical representation of the results, it is immediate to conclude that only one ---Eq.~\eqref{ann}--- of the three restrictions presented in this section is required to constrain this parameter space because the two straight lines ---Eq.~\eqref{antisym_border_restriction}--- are tangent to the blue parabola ---Eq.~\eqref{ann}.

Similarly to the symmetric subcase, in this study, the minimum of the potential might not be reached when considering the minimum values of $\rho^{\si{eff}}_{i}\ i=3,6,7$, and a new notation with the subscript $V_{\si{min}}$ is introduced. It is important to highlight that, to the best of our knowledge, similar results to those presented in Equations \eqref{antisym_border_restriction} and \eqref{ann} have not yet been derived, either analytically or numerically.

\subsubsection{General 2HDM potential}\label{generalll}

Lastly, we aim to study the most general 2HDM potential. In this case, the angular part of the potential is
\begin{equation}\label{new_hs2}
f(\chi)=2+\left(\rho^{\si{eff}}_{3}-1\right)\sin^2{\left(2\chi\right)}+u\sin{\left(2\chi\right)}+v\sin{\left(2\chi\right)}\cos{\left(2\chi\right)}\ .
\end{equation}
The conditions in this particular situation can be briefly expressed as:
\begin{equation}\label{generalpot_conclusions}
    \inf\limits_{\chi\in (0,\frac{\pi}{2})}{f(\chi\ ;\rho^{\si{eff}}_{3,\si{V}_{\si{min}}},u_{\si{V}_{\si{min}}},v_{\si{V}_{\si{min}}})}>0\ .
\end{equation}

\begin{figure}[!t]
    \centering
    \includegraphics[scale=0.25]{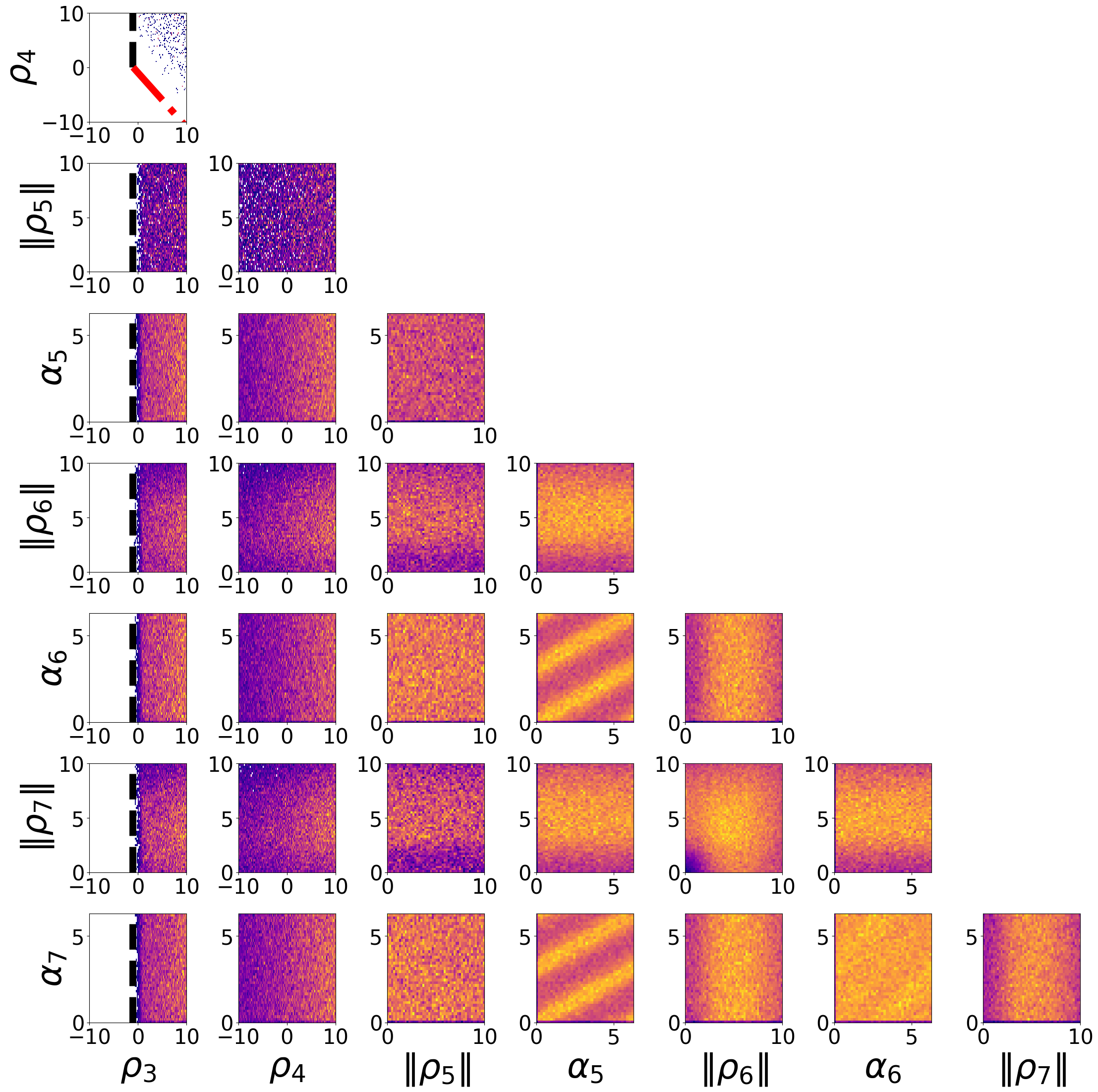} 
    \caption{Valid parameters of the most general $\mathbb{Z}_2$-violating 2HDM potential have been calculated computationally \cite{javi}. The dashed black line corresponds to $\rho_{3}=-1$, while the dash-dotted red line, to the restriction $\rho_{3}=-1-\rho_{4}$.}
    \label{stairgeneralplot}
\end{figure}

However, in contrast with the symmetric and antisymmetric subcases, finding all closed-form expressions for the restrictions in the general case can be highly involved. 
Because of this, and in light of the previous analyses for the simpler cases, a numerical approach will be considered and the values of $\rho^{\si{eff}}_{3}$, $u$ and $v$ that minimize the potential \label{new_hs} will be obtained in our public Rust code~\cite{javi}. Nevertheless, note that a simple but strong condition can be derived easily. Since both Higgs doublets can explore the entire region $r\in[0,1]$ ---$r$ is a free parameter--- one possible value is $r=0$. In this scenario, Equation~\eqref{new_hs2} reduces to
\begin{equation}
    2 + \left(\rho_{3}-1\right)\sin^2{\left(2\chi\right)} > 0.
\end{equation}
In other words, we recover Equation~\eqref{sameingeneralpot} and the same reasoning that led us to Equation~\eqref{conditionsinglet} can be applied in this case. This means that this result is true in the most general potential and, therefore, in all the subcases that have been presented herein. This mathematical result can be corroborated computationally in Figure~\ref{stairgeneralplot}, where one can observe all the possible projections of the parameter space. Another interesting result that is shown in that figure is that although the restriction $\rho_{3} + \rho_{4} = -1$ ---dash-dotted red line--- has been derived previously when studying the $\mathbb{Z}_2$-conserving 2HDM, it seems to be also valid when the $\mathbb{Z}_2$ symmetry is hardly broken. Furthermore, in view of Figure~\ref{stairgeneralplot}, besides the dotted lines, there may not be additional constraints which have not yet been derived mathematically. The fact that led us to this conclusion is that it seems to exist a homogeneous distribution of points in every subplot.

We also observe that although the whole phase-space is allowed for the phase-shifts $\alpha_5$, $\alpha_6$ and $\alpha_7$, Fig.~\ref{stairgeneralplot} shows the presence of important correlations, favoring configurations with $\alpha_5\approx 2 \alpha_6$ and $\alpha_5 \approx 2(\alpha_7+\pi)$.
Note that the projection onto the $\rho_{3}-\rho_{4}$ plane appears less dense than the others because many of the points shown in the remaining projections lie outside the plotting range, however this projection may require further analysis.

\begin{figure}[!t]
    \centering
    \includegraphics[scale=0.2]{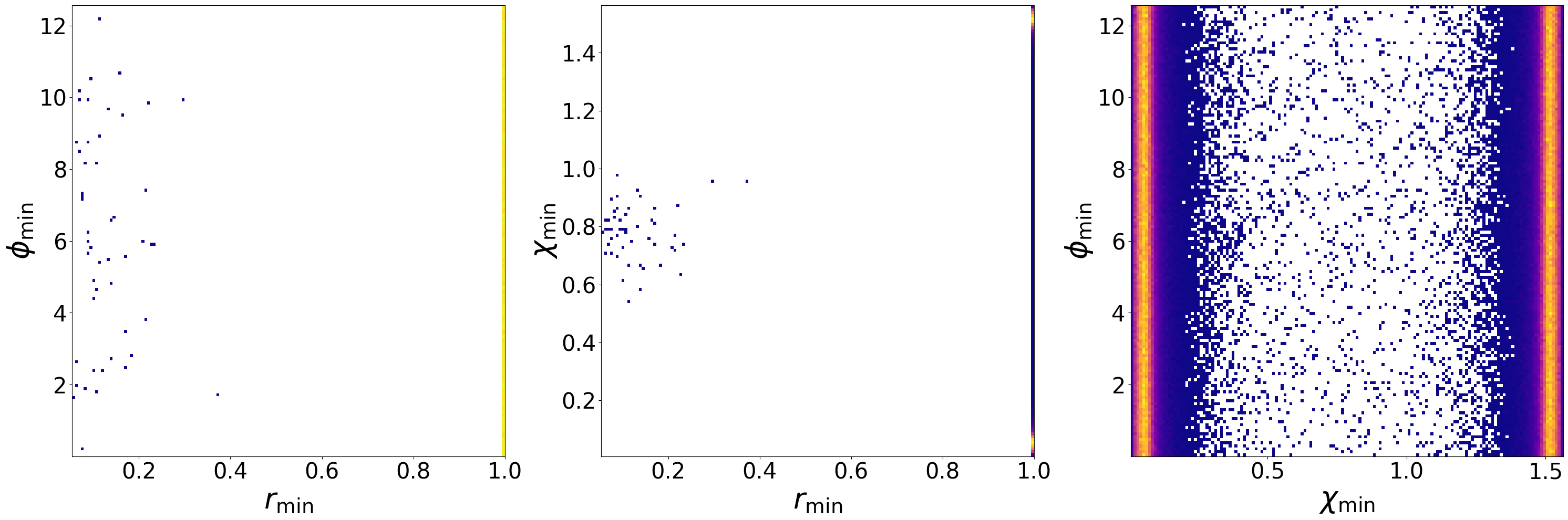} 
    \caption{Projections of the minimization space. The vast majority of parameters ---which were randomly generated--- minimize in $r=1$ and in two $\chi$ values near the borders. The distribution in the $\phi$ variable seems to be uniform.}
    \label{minimization_points}
\end{figure}

\begin{figure}[!t]
    \centering
    \includegraphics[scale=0.25]{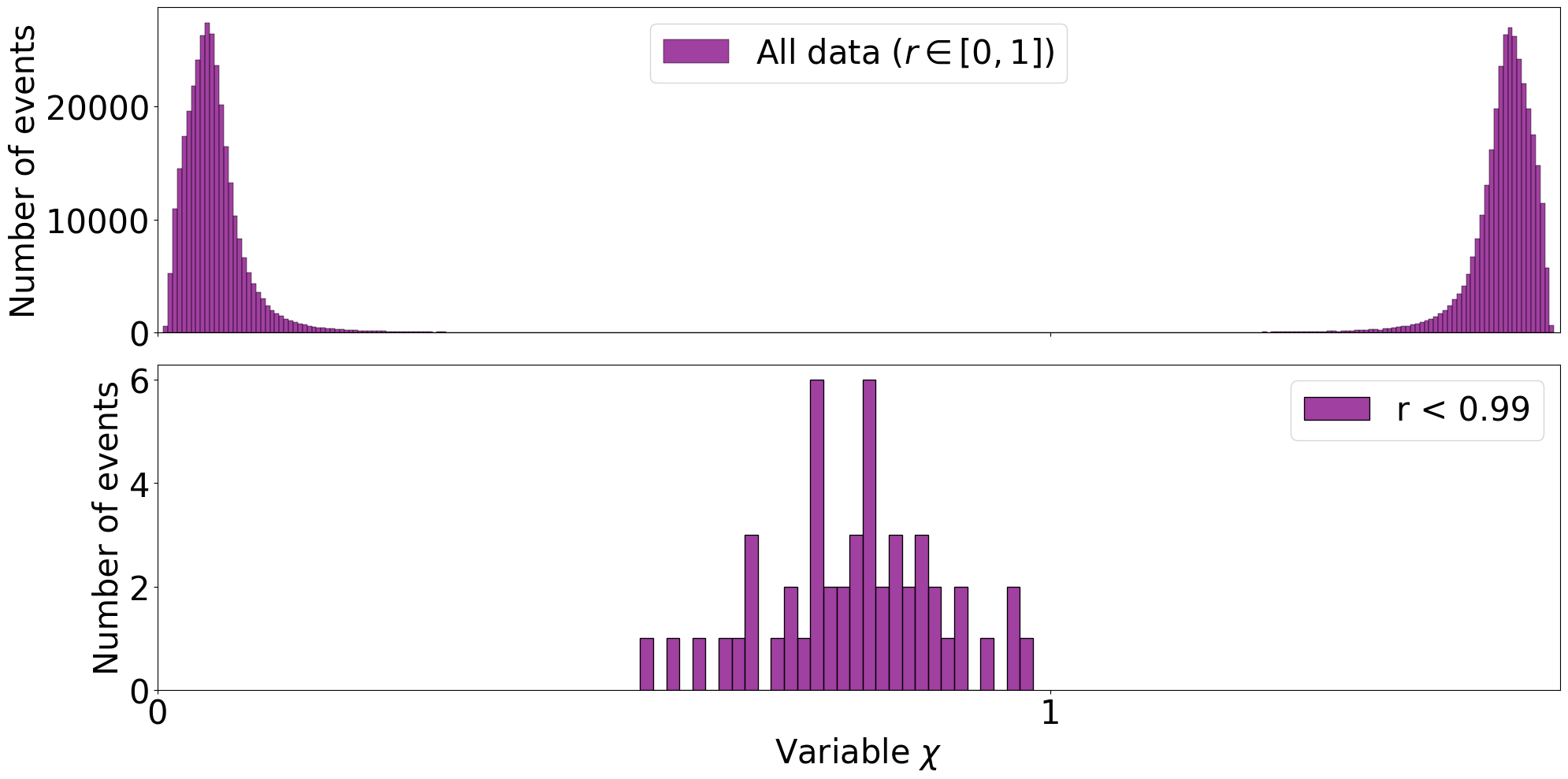} 
    \caption{Number of events in the $\chi$ variable. Quantitative analysis suggests that given a randomly generated set of parameters, they tend to minimize in $r=1$ and in two $\chi$ values near the borders. However, when both Higgs doublets are perpendicular ($r=0$), the potential minimizes near the center of the first quadrant, $\chi=\pi/4$.  }
    \label{histplot}
\end{figure}

Another important result is where the minimization occurs in $(r,\phi,\chi)$. In Fig.~\ref{minimization_points} and Fig.~\ref{histplot}, it is possible to observe that the major number of events minimize in $r=1$, where both Higgs doublets are parallel. In fact, in this subcase, there exist two ``sweet'' angles that are near to the borders of the orthant where the minimization tends to take place. 
On the contrary, when the minimization occurs near $r=0$ (see Fig.~\ref{minimization_points}) ---the subcase where both doublets are perpendicular ($n_1\perp n_2$)--- the result also tends to minimize near $\chi=\pi/4$, in the exact middle of the first quadrant.

 A final afterthought can be extracted here: our statistical evaluation of the most general 2HDM quartic potential reveals a strong tendency in the parameter space to get minimized along these two specific ``sweet'' angles. These directions in the field space run close the boundaries of the orthant (i.e., along values $\Phi_1^\dagger \Phi_1\ll \Phi_2^\dagger \Phi_2$, or vice versa). It is crucial to distinguish this angular minimization ---which strictly enforces global copositivity and boundedness from below in the asymptotic large-field limit--- from the minimization procedure that determines the physical vacuum. Geometrically, these preferred angles parametrize the trajectories in field space along which the potential exhibits minimal growth as the scalar field moduli tend to infinity. Nonetheless, it is compelling to consider that these asymptotic directions of minimal quartic growth might inherently shape the vacuum topology, leaving an imprint on the hierarchies of the two fields for their vacuum expectation values (vev) and masses.

\section{2HDM softly-broken \texorpdfstring{$\mathbb{Z}_2$}{Z2} phenomenology: relation between the \texorpdfstring{$\rho_{i}$}{rhoi} and physical parameters}
\label{sectionIV}
\label{sec:pheno-2HDM-softly-broken}

To assess the impact of these copositivity bounds on physical parameters, this section focuses on a softly-broken $\mathbb{Z}_2$ model. Here, the quartic sector preserves the $\mathbb{Z}_2$ symmetry, whereas quadratic terms are permitted to break it softly~\cite{Dawson_2022,Dawson:2023oce}.
This softly-broken 2HDM scenario has been widely applied both to collider phenomenology~\cite{Dawson_2022,Dawson:2023oce,Arco:2022jrt,Arco:2023sac,Matuszak:2026xsz,Moretti:2026qsw} and cosmology analyses~\cite{Matuszak:2026xsz,Moretti:2026qsw}, what makes the discussion in this Section timely and relevant.

The constraints obtained in previous sections are expressed in terms of the parameters of the potential. However, these parameters are not directly measurable in experiments or have immediate physical interpretations. Because of this, it is convenient to rewrite them in terms of physical quantities such as scalar masses and mixing angles. 

We will derive explicit expressions for $\rho_{i}$ ($i=3,4,5$) in terms of the relevant physical parameters, in order to analyze how our positivity constraints, based on the idea presented by 
Lemma~\ref{comment:lemma-suff} and
Lemma~\ref{comment:lemma-necessary}, restrict the allowed region of the physical parameter space. We will not consider the restrictions derived from the photon mass of Eq.~\eqref{vacionnocargado} for this study so as to be able to compare the necessary conditions 
from Lemma \ref{comment:lemma-necessary} to the sufficient conditions 
from Lemma~\ref{comment:lemma-suff} that arise from a positive definite matrix $\Lambda$. For simplicity of the analysis, the study will be carried out only on the softly-broken $\mathbb{Z}_2$ 2HDM potential.

In order to express $\rho_{i}$ in terms of the physical parameters, we first consider a rotation of the entire $\mathbb{Z}_2$-conserving 2HDM lagrangian to the Higgs basis \cite{CilleroQuezadaHiggsbasis}:
\begin{equation}
    \begin{pmatrix}
        H_1\\
        H_2
    \end{pmatrix}=\underbrace{\begin{pmatrix}
        \cos\beta & \sin\beta\\
        -\sin\beta & \cos\beta
    \end{pmatrix}}_{R(\beta)}\begin{pmatrix}
        \Phi_1\\
        \Phi_2
    \end{pmatrix}\ ,
\end{equation}

where $\tan\beta=v_2/v_1$, and $v_1$, $v_2$ are the vacuum expectation values   
of $\Phi_{1}$ and $\Phi_{2}$, respectively. By virtue of the foregoing, it is possible to determine how all combinations of $\Phi^{\dagger}_{a}\Phi_{b}$, with $a,b=1,2$, transform: 
\begin{equation}
    \begin{pmatrix}
        H^{\dagger}_{1}H_{1}\\ H^{\dagger}_{2}H_{2}  \\ H^{\dagger}_{1}H_{2}\\ H^{\dagger}_{2}H_{1}
    \end{pmatrix}=\mathcal{Q}^{-1}(\beta)\begin{pmatrix}
        \Phi^{\dagger}_{1}\Phi_{1}\\  \Phi^{\dagger}_{2}\Phi_{2}     \\ \Phi^{\dagger}_{1}\Phi_{2}\\ \Phi^{\dagger}_{2}\Phi_{1}
    \end{pmatrix}\ .
\end{equation}

The quartic terms of the potential preserve \texorpdfstring{$\mathbb{Z}_2$}{Z2} symmetry and can be written by means of a quadratic form, $\Lambda^{mn}$,   
\begin{equation}\label{pot2hdmcompacto}
    V^{\backslash\hspace*{-0.15cm}\mathbb{Z}_2\ (4)}_{2\si{HDM}}= 
    \frac{1}{2} \begin{pmatrix} 
    \Phi^{\dagger}_{1}\Phi_{1},\Phi^{\dagger}_{2}\Phi_{2} ,\Phi^{\dagger}_{2}\Phi_{1},\Phi^{\dagger}_{1}\Phi_{2}\end{pmatrix}
    \, \Lambda \,       
    \begin{pmatrix}
        \Phi^{\dagger}_{1}\Phi_{1}\\ \Phi^{\dagger}_{2}\Phi_{2} \\ \Phi^{\dagger}_{1}\Phi_{2}\\ \Phi^{\dagger}_{2}\Phi_{1}     \end{pmatrix}\,,
     \quad \mbox{with }  
    \Lambda\,=\, \begin{pmatrix}
        \lambda_1 & \lambda_3 & 0 &0 \\ \lambda_3 & \lambda_2 &0&0  \\ 0&0&\lambda_4&\lambda_5^*  \\ 0&0&\lambda_5&\lambda_4\end{pmatrix}  \, , 
\end{equation}
where $\Lambda$ has eigenvalues $\frac{1}{2}\left(\lambda_1+\lambda_2 \pm \sqrt{(\lambda_1-\lambda_2)^2 + 4 \lambda_3^2}\right)$ and $\lambda_4\pm ||\lambda_5||$. The naive eigenvalue positivity requirement on this quadratic form $\Lambda$ leads to the ``sufficient'' constraints (according to Lemma~\ref{comment:lemma-suff}) $\lambda_1,\, \lambda_2 >0 $ and $\lambda_1\lambda_2-\lambda_3^2>0$ (referred in Sec.~\ref{sec:RSE}), 
and the additional condition $||\lambda_5||< \lambda_4$.

By means of the previous relations 
between $\Phi^{\dagger}_{a}\Phi_{b}$ and $H^{\dagger}_{a}H_{b}$, it is possible to find a relation between $Z_{i}$ ---the parameters of the potential when written in the Higgs basis--- 
 and $\lambda_{i}$~\cite{Branco_2012},   
\begin{equation}
    \mathcal{Z}=\mathcal{Q}^{\dagger}(\beta)\Lambda\mathcal{Q}(\beta)\ , 
\end{equation}
where $\mathcal{Z}^{mn}$ contains all $Z_{i}$. Once it is known how to express $\lambda_{i}$ in terms of $Z_{i}$, one can introduce a previously calculated well-known relation \cite{Dawson_2022, CilleroQuezadaHiggsbasis} between $Z_{i}$ and the more physical parameters: $Y_{2}$, the boson masses, the vev ratio $\beta$ and an extra rotation parameter, $\alpha$. Additional degrees of freedom disappear when imposing the equations that arise from minimizing the potential when calculating the vev. In the end, we have exchanged the $\rho_i$ parameters of the Lagrangian for the following parameters:
\begin{equation}
    \{\tan \beta, v,m_h, M_H, Y_2,\cos (\beta-\alpha)\}\ .
\end{equation}

\begin{figure}[!t]
    \centering
    \includegraphics[scale=0.24]{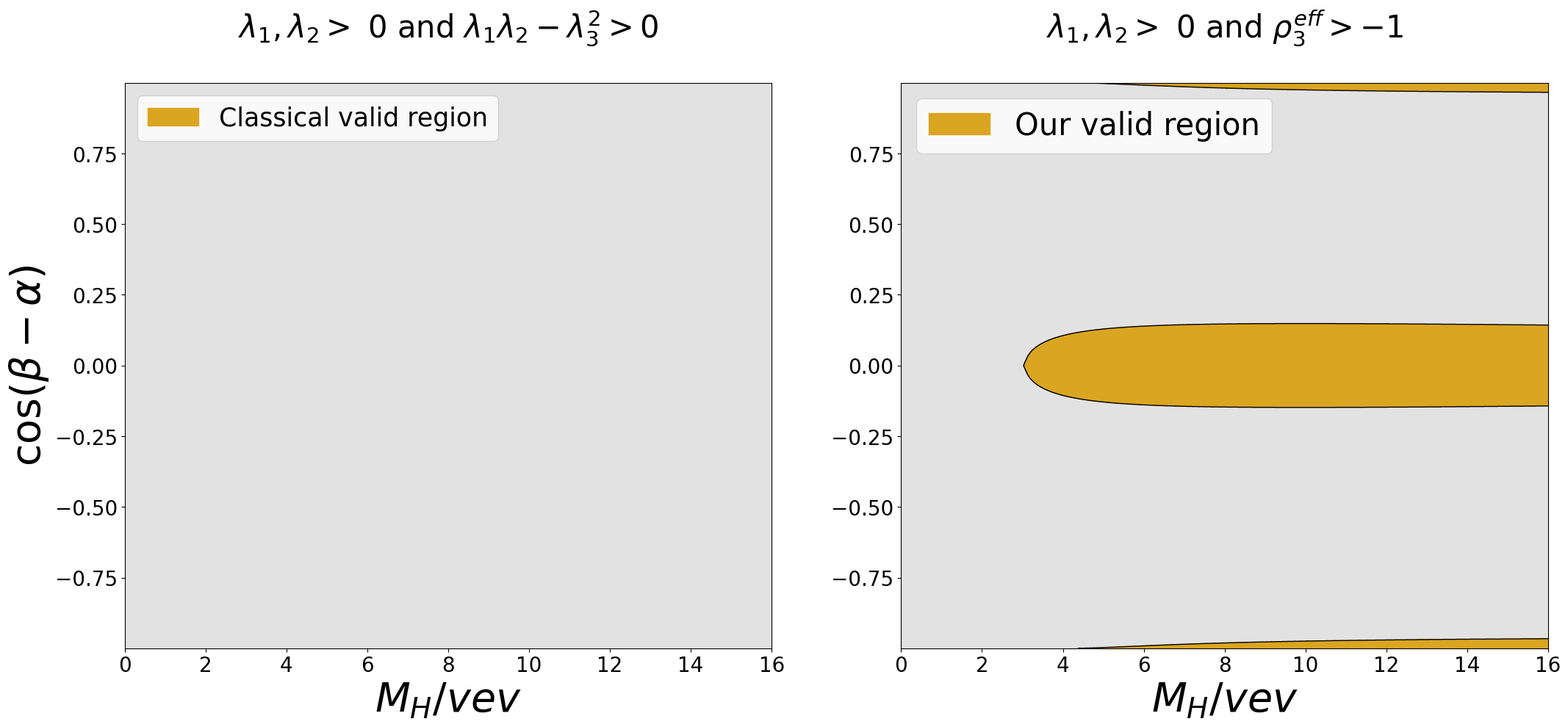} 
    \caption{The parameters used in this plot are $m_{h} = 125\ \si{GeV}$, $Y_{2} = \left(750\ \si{GeV}\right)^2$, $\tan{\beta}=v_2/v_1=1$ and vev $= \sqrt{v^{2}_{1}+v^{2}_{2}} = 246\ \si{GeV}$. \textit{Left}: Among the restrictions that are necessary to have a positive definite matrix $\Lambda$ in this softly-broken $\mathbb{Z}_{2}$ scenario, only $\lambda_{1}\lambda_{2}-\lambda^2_{3}>0$ and $\lambda_{1}, \lambda_{2} > 0$ have been considered because they forbid all the plotted region, adding more restrictions will only lead us to the same scenario. \textit{Right}: The conditions obtained in Subsection \ref{subsect_softlybrokenZ2} constrain the physical parameter space.}
    \label{Restrictioncomparison}
\end{figure}

\begin{figure}[!t]
    \centering
    \includegraphics[scale=0.45]{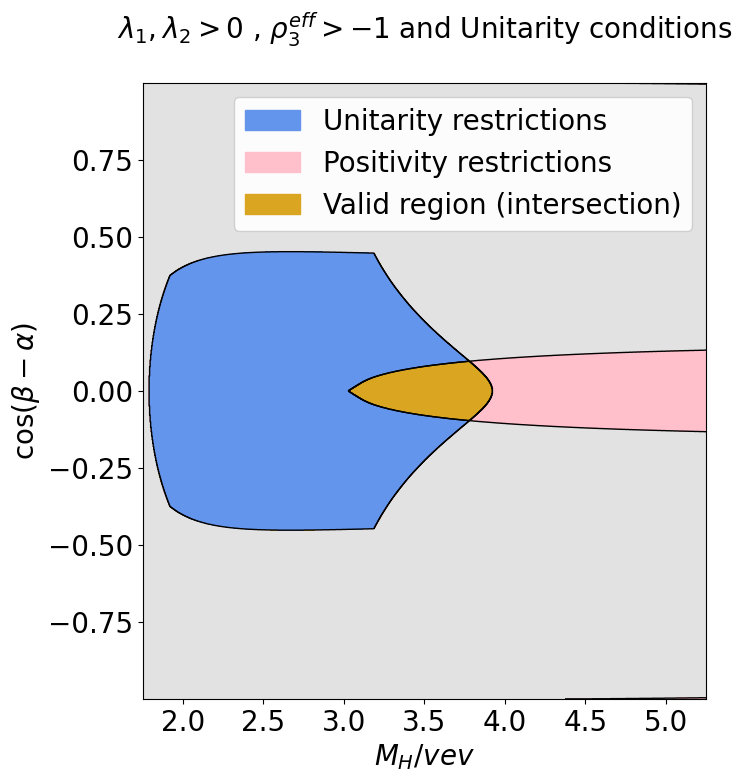} 
    \caption{Positivity and unitarity conditions constrain simultaneously the parameter space. The mass of Higgs boson of the Standard Model has been taken as $m_{h} = 125\ \si{GeV}$ and the masses of the rest of the particles have been considered to be equal to the mass of the heavy Higgs, which is a free parameter, $M_{H}$. In addition, $Y_{2} = (750\ \si{GeV})^2 $, $\tan{\beta}=v_2/v_1=1$ and vev= $\sqrt{v_1^2+v_2^2}= 246\ \si{GeV}$ . Note that the heavy Higgs mass is bounded between $740\ \si{GeV}$ and $990\ \si{GeV}$ approximately ---three to four times the vacuum expectation value.  }
    \label{Unitarity_and_positivity_restriction}
\end{figure}
Now, it is possible to translate the restrictions on $\rho_{i}$ to the physical parameters. This has been done in Figure~\ref{Restrictioncomparison} and one can see that, as anticipated, the sufficient conditions of Lemma~\ref{comment:lemma-suff} are excessively restrictive \footnote{It is important to note that, in this figure, for the sake of simplicity, the mass of Higgs boson of the Standard Model, $m_{h}$, has been fixed to $125\ \si{GeV}$, while the masses of the rest of the particles have been considered to be equal to the mass of the heavy Higgs, which is a free parameter, $M_{H}$. }. Indeed, given this usual positivity constraint, the valid region ---herein indicated as ``classical valid region''--- disappears and the 2HDM theory would instantly become untenable. Fortunately, the necessary conditions presented in Lemma~\ref{comment:lemma-necessary} allow a small physical region. From a copositivity perspective, this does not preclude previous investigations and leaves room for future analyses. 

Another relevant restriction to apply to the potential are the unitarity conditions \cite{ginzburg2003treelevelunitarityconstraints2hdm,ginzburg2005symmetries2hdmdifferentvacua,Ginzburg_2005, Ginzburg_2005_2}, see Figure~\ref{Unitarity_and_positivity_restriction}. This restriction merely implies that certain non-linear combinations of $\lambda_{i}$, commonly referred to as $\Lambda_{Y\sigma\pm}$, must be much less than $8\pi$, that is, $|\Lambda_{Y\sigma\pm}|\ll 8\pi$. However, following a line of reasoning similar to that found in the literature \cite{ginzburg2003treelevelunitarityconstraints2hdm,ginzburg2005symmetries2hdmdifferentvacua,Ginzburg_2005, Ginzburg_2005_2}, we will consider the limit where $|\Lambda_{Y\sigma\pm}| < 8\pi$. Here, we adopt the same notation as used in the references.

In Figure~\ref{Unitarity_and_positivity_restriction}, one can see that the positivity restrictions constitute a lower bound for the heavy Higgs mass, whilst the unitarity conditions bound these physical parameters from above. These results show that the heavy Higgs mass is constrained between $740\ \si{GeV}$ and $990\ \si{GeV}$ approximately ---three to four times the vacuum expectation value (vev $\approx 246\ \si{GeV}$). 
Although this TeV energy range is currently accessible and has been thoroughly probed by recent high-energy collision runs at the LHC~\cite{ATLAS:2024eqt, CMS:2024vxp}, the absence of any confirmed experimental evidence for a second Higgs boson does not fully preclude the possibility of living in a universe governed by a 2HDM theory.

\section{Conclusions}
\label{sec:conclusions}

Higgs extensions have been extensively studied in the literature: from an additional scalar field \cite{Buchalla_2017,Robens_2015,Kannike_2016,kannike2024rebuttalnotevacuumstability,BISWAS_2013,Barger_2009,Cho_2023} to more complex $N$-Higgs models, of which 2HDM \cite{Branco_2012,Ginzburg_2005, Ginzburg_2005_2,ginzburg2005symmetries2hdmdifferentvacua,Gunion_2003,kastening1993boundsstabilitysymmetrybreaking,Deshpande:1977rw,Dawson_2022,Baum_2018,CilleroQuezadaHiggsbasis,gonzalez-jerez-quezada-sanz-cillero-2017,Boto_2020, ginzburg2003treelevelunitarityconstraints2hdm} is a special case. The reason behind imposing theoretical constraints to a certain Higgs extension ---one that might better describe nature than the current Standard Model does--- is because it may lead us to a small, bounded parameter space on which we shall focus our experimental attention. In this work an interesting variety of different Higgs extensions have been studied in detail. 

First we studied the RSE extension and recovered the conditions already obtained for the strictest and the most relaxed constraints. In order to do this, we parametrized the bilinear of quadratic fields by considering the change of variables given by Eq.~\eqref{first_changeofvariable} and studied the relevant orthant. As expected, the more restrictive conditions require $\rho_{3}$ in the validity range to $(-1,1)$, while the less restrictive conditions are more permissive as they allow $\rho_{3}\in (-1,\infty)$.

Then we moved onto the 2HDM, we considered a similar change of variables given by Eq.~\eqref{2HDMvariables}. This allowed us to translate the restrictions obtained in Section \ref{section2} after imposing positivity ---which were extremely straightforward to derive--- to a similar constraint in Section \ref{section3} for the 2HDM, where the potential was considerably more involved. We first explored the $\mathbb{Z}_2$ conserving 2HDM and used the same technique finding an agreement with previous works. Additionally, we developed a Rust code~\cite{javi} capable of finding these restrictions numerically. As we can see in Fig.~\ref{Z2_projection_rho_345}, the parameter space complies with the theoretical restrictions found in Eqs.~\eqref{real_cond} and \eqref{cond_over_NO_effective_rho3}. 

To address the most general 2HDM including $\mathbb{Z}_2$ violation, we divided this scenario into three subcases regarding the values of $\rho_6$ and $\rho_7$.  
We verified the analytical restrictions for the first two subcases using a custom \textit{Rust} implementation, which generated Figs.~\ref{sym_restiction} and \ref{antisym_restriction} in full agreement with the bounds in Eqs.~\eqref{minimo_hs}, \eqref{antisym_border_restriction}, and \eqref{ann}. Furthermore, for the general 2HDM scenario, the numerical output shown in Fig.~\ref{stairgeneralplot} complies with the restrictions of Eqs.~\eqref{real_cond} and \eqref{cond_over_NO_effective_rho3}. Interestingly, despite explicit $\mathbb{Z}_2$ symmetry violation in this potential, the numerical results closely follow the bounds derived for the $\mathbb{Z}_2$-conserving Lagrangian in those two Equations.  
Finally, the homogeneous distribution of points across the valid regions in Fig.~\ref{stairgeneralplot} suggests the absence of additional, unmapped constraints.

Besides the mathematical and numerical constraints, one of the initial motivations behind this work was also to ascertain that the classical positivity restrictions ---those which arise when imposing the matrix $\Lambda$ to be positive definite, see Equations \eqref{quartictermssinglet} and \eqref{pot2hdmcompacto}--- excessively constrain the parameter space. By virtue of Figures~\ref{DSHM} and \ref{Restrictioncomparison}, one can immediately corroborate this fact. Indeed, in the second figure, it is possible to see that our conditions allow for small and extreme values of $\cos{\left(\beta - \alpha\right)}$. This region, added to the unitarity constraints \cite{ginzburg2003treelevelunitarityconstraints2hdm,ginzburg2005symmetries2hdmdifferentvacua,Ginzburg_2005, Ginzburg_2005_2} shown in Figure~\ref{Unitarity_and_positivity_restriction}, bounds the heavy Higgs mass between $740\ \si{GeV}$ and $990\ \si{GeV}$ approximately.

To finish, it is crucial to remember that all the positivity conditions that have been presented throughout this work are based on the assumption that the dominant terms of the studied potentials are always the quartic terms. 
However, as shown in Figure~\ref{Quadratic_terms_2HDM} in App.~\ref{quadratictermsandloops}, there are particular flat directions in field space where this assumption no longer holds, making the lower-order terms essential for determining the overall stability of the potential. See App.~\ref{quadratictermsandloops} for a detailed analytical discussion of this specific subcase.  

In summary, replacing excessively restrictive matrix positivity conditions with exact copositivity criteria significantly expands the theoretically viable parameter space of extended Higgs sectors. By combining these refined analytical constraints with numerical tools, we achieve a more faithful mapping with wider allowed regions. This broader parameter space is crucial for guiding phenomenological studies and experimental searches.

\section*{Acknowledgements}

C. Quezada-Calonge acknowledges financial support from the Università di Torino through the DataSMEFT23 grant (PNRR – NextGenerationEU) and from the Istituto Nazionale di Fisica Nucleare (INFN) through the Iniziativa Specifica SPIF. 
This work is partially supported by the projects
PID2025-170350NB-I00 and PID2022-137003NB-I00
funded by MICIU/AEI/10.13039/ 501100011033 and by ERDF/EU, and Universidad Complutense de Madrid under research group 910309 and
the IPARCOS institute.

\begin{appendices}
 
\section{Quadratic terms}\label{quadratictermsandloops}

In all the positivity restrictions analyzed in this work, we have used strict inequalities in order to exclude the case where the potential exhibits a certain direction in the first quadrant along which it remains constant ---it neither increases nor decreases. Nevertheless, there are some paramount aspects that have not yet been taken into consideration that can allow the use of non-strict inequalities. In order to show this, we will study the $\mathbb{Z}_2$-conserving 2HDM as an example; this analysis can also be extended to other models.

\subsection{2HDM with \texorpdfstring{${Z}_2$}{Z2} symmetry}

In this Higgs extension, it is undoubted that Equation~\eqref{real_cond} bounds the potential of Eq.~\eqref{2HDMpotential_newvariables} from below. Moreover, as it was shown before, considering the limit where $\rho^{\si{eff}}_{3,\si{min}} = -1$ was not acceptable at first glance. However, the possibility of using a non-strict inequality arises when other terms ---instead of the quartic terms that have been studied--- become dominant. Note that despite the quadratic terms are always subdominant ---compared to quartic terms--- in the limit we are considering, one can verify that quartic terms disappear in the line $y=x$ when $\rho^{\si{eff}}_{3}=-1$. As a consequence, quadratic terms acquire importance in this situation. If quadratic terms are not disregarded, the entire $\mathbb{Z}_2$-conserving 2HDM potential can be written as:
\begin{equation}\label{2HDM_entero_new_variables}
    V^{\mathbb{Z}_2}_{2\si{HDM}}=-\sigma_{1}x-\sigma_{2}y+\frac{1}{2}x^2+\frac{1}{2}y^2+\rho^{\si{eff}}_{3}xy\ ,
\end{equation}
where
\begin{equation}
    \sigma_{i}\equiv\frac{m^2_{ii}}{2\sqrt{\lambda_{i}}}\ ,\ i=1,2.
\end{equation}

In view of Equation~\eqref{2HDM_entero_new_variables}, taking $\rho^{\si{eff}}_{3}=-1$ and $y=x$ the potential reduces to:
\begin{equation}
V^{\mathbb{Z}_2}_{2\si{HDM}} \Big|_{\substack{\rho^{\text{eff}}_{3}=-1 \\ y=x}} = -(\sigma_{1}+\sigma_{2})x\ .
\end{equation}

\begin{figure}[!t]
    \centering
    \includegraphics[scale=0.4]{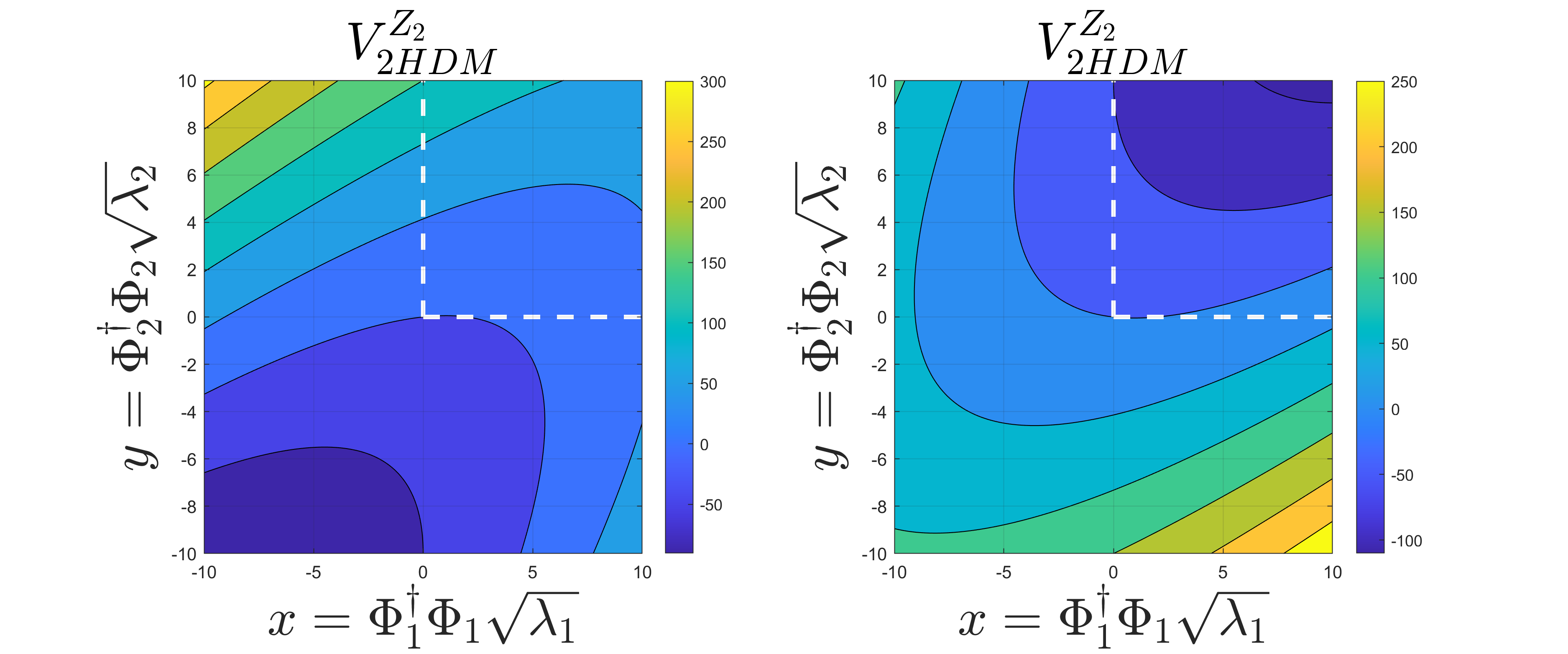} 
    \caption{$\mathbb{Z}_2$-conserving 2HDM potential with $\rho^{\si{eff}}_{3}=-1$. \textit{Left}: $\left(\sigma_{1},\sigma_{2}\right)=\left(1,-10\right)$. \textit{Right}: $\left(\sigma_{1},\sigma_{2}\right)=\left(1,10\right)$.}
    \label{Quadratic_terms_2HDM}
\end{figure}

In order to bound the potential from below, it must increase for each possible direction in the first quadrant. Hence, one can impose that
\begin{equation}\label{finalresult_2HDM_Z2}
    \rho^{\si{eff}}_{3,\si{min}}\geq -1\ ,
\end{equation}
provided $\sigma_{1}+\sigma_{2}<0$ when $\rho^{\si{eff}}_{3,\si{min}}=-1$, see Figure~\ref{Quadratic_terms_2HDM}.

\subsection{Other Higgs extensions}

In other models, the restrictions may vary considerably after studying the limit cases. For instance, in a $\mathbb{Z}_2$-violating potential, the next terms that need to be considered after quartic terms are cubic combinations of the fields, instead of quadratic terms. Additionally, other factors ---which have not been considered within this work--- can alter the restrictions obtained only considering the highest order terms. As an example, loops might perturb the potential to the extent that these corrections are comparable to quartic terms and, therefore, it would result in a shift of the lower bound for $\rho^{\si{eff}}_{3,\si{min}}$. In any case, these general corrections exceed the scope of this paper and would be interesting to study in a future work.

\end{appendices}

\bibliographystyle{SciPost_bibstyle}
\bibliography{references}

\end{document}